\documentclass[article, prl, nobibnotes, secnumarabic, amssymb, superscriptaddress, twocolumn, aps,longbibliography]{revtex4-2}

\usepackage{chemformula}
\usepackage{upgreek}
\usepackage{bm}
\usepackage{mathrsfs}
\usepackage{hyperref}
\usepackage{xcolor}
\usepackage{url}
\hypersetup{colorlinks,breaklinks,
            urlcolor=[rgb]{0,0,0.64},
            linkcolor=[rgb]{0,0,0.64},
            citecolor=[rgb]{0,0,0.64},
            filecolor=[rgb]{0,0,0.64}}

\begin{document}

\newcommand{\tns}{Ta$_\text{2}$NiSe$_\text{5}$}
\newcommand{\tc}{$T_{\text{c}}$\xspace}
\newcommand{\mjc}{mJ/cm$^2$\ }
\newcommand{\wn}{cm$^{-1}$\ }

\newcommand{\LLP}{Key Laboratory for Laser Plasmas (Ministry of Education), School of Physics and Astronomy, Shanghai Jiao Tong University, Shanghai 200240, China.}
\newcommand{\TDL}{Tsung-Dao Lee Institute, Shanghai Jiao Tong University, Shanghai 201210, China.}
\newcommand{\IFSA}{Collaborative Innovation Center of IFSA (CICIFSA), Shanghai Jiao Tong University, Shanghai 201210, China.}
\newcommand{\ZIAS}{Zhangjiang Institute for Advanced Study, Shanghai Jiao Tong University, Shanghai 201210, China.}
\newcommand{
\LASQC}{Key Laboratory of Artificial Structures and Quantum Control (Ministry of Education), School of Physics and Astronomy, Shanghai Jiao Tong University, Shanghai 200240, China.}
\newcommand{\SHTU}{School of Physical Science and Technology, ShanghaiTech University, Shanghai 201210, China.}
\newcommand{\CU}{Department of Physics and Astronomy, Clemson University, Clemson, South Carolina 29634, U.S.A.}
\newcommand{\UCB}{Physics Department, University of California, Berkeley, California 94720, U.S.A.}
\newcommand{\UCBC}{Department of Chemistry, University of California, Berkeley, California 94720, U.S.A.}
\newcommand{\LBNL}{Materials Science Division, Lawrence Berkeley National Lab, Berkeley, California 94720, U.S.A.}
\newcommand{\StanfordAP}{Departments of Physics and of Applied Physics, Stanford University, Stanford, California 94305, U.S.A.}
\newcommand{\SIMES}{Stanford Institute for Materials and Energy Sciences, SLAC National Accelerator Laboratory, Menlo Park, California 94025, U.S.A.}
\newcommand{\PKU}{International Center for Quantum Materials, School of Physics, Peking University, Beijing 100871, China.}
\newcommand{\SLAB}{Songshan Lake Materials Laboratory, Dongguan, Guangdong, China.}
\newcommand{\CICAM}{Collaborative Innovation Center of Advanced Microstructures, Nanjing University, Nanjing 210093, China.}
\newcommand{\CAS}{Beijing National Laboratory for Condensed Matter Physics, Institute of Physics, Chinese Academy of Sciences, Beijing 100190, China.}
\newcommand{\BIT}{Centre for Quantum Physics, Key Laboratory of Advanced Optoelectronic Quantum Architecture and Measurement (MOE), School of Physics, Beijing Institute of Technology, Beijing 100081, China.}
\newcommand{\BITN}{Beijing Key Lab of Nanophotonics and Ultrafine Optoelectronic Systems, Beijing Institute of Technology, Beijing 100081, China.}
\newcommand{\BITZ}{Beijing Institute of Technology, Zhuhai 519000, China.}
\newcommand{\BAQIS}{Beijing Academy of Quantum Information Sciences, Beijing 100913, China.}


\title{Structural contribution to light-induced gap suppression in Ta$_2$NiSe$_5$}

\author{Zijing Chen}
\thanks{These authors contributed equally to this work: Zijing~Chen, Chenhang~Xu, and Chendi~Xie.}
\affiliation{\LLP}
\affiliation{\IFSA}
\affiliation{\ZIAS}

\author{Chenhang Xu}
\thanks{These authors contributed equally to this work: Zijing~Chen, Chenhang~Xu, and Chendi~Xie.}
\affiliation{\StanfordAP}
\affiliation{\SIMES}
\author{Chendi Xie}
\thanks{These authors contributed equally to this work: Zijing~Chen, Chenhang~Xu, and Chendi~Xie.}
\affiliation{\CU}

\author{Weichen Tang}
\affiliation{\UCB}
\affiliation{\LBNL}

\author{Qiaomei Liu}
\affiliation{\PKU}

\author{Dong Wu}
\affiliation{\BAQIS}

\author{Qing Xu}
\affiliation{\LLP}
\affiliation{\IFSA}
\affiliation{\ZIAS}

\author{Tao Jiang}
\affiliation{\LLP}
\affiliation{\IFSA}
\affiliation{\ZIAS}

\author{Pengfei Zhu}
\affiliation{\LLP}
\affiliation{\IFSA}
\affiliation{\ZIAS}
\affiliation{\TDL}

\author{Xiao Zou}
\affiliation{\LLP}
\affiliation{\IFSA}
\affiliation{\ZIAS}

\author{Jun Li}
\affiliation{\CAS}

\author{Zhiwei Wang}
\affiliation{\BIT}
\affiliation{\BITN}
\affiliation{\BITZ}

\author{Nanlin Wang}
\affiliation{\PKU}
\affiliation{\BAQIS}

\author{Dong Qian}
\email[Correspondence to: ]{dqian@sjtu.edu.cn}
\affiliation{\TDL}
\affiliation{\CICAM}
\affiliation{\LASQC}

\author{Alfred Zong}
\email[Correspondence to: ]{alfredz@stanford.edu}
\affiliation{\StanfordAP}
\affiliation{\SIMES}

\author{Dao Xiang}
\email[Correspondence to: ]{dxiang@sjtu.edu.cn}
\affiliation{\LLP}
\affiliation{\IFSA}
\affiliation{\ZIAS}
\affiliation{\TDL}

\date{\today}

\begin{abstract}
An excitonic insulator is a material that hosts an exotic ground state, where an energy gap opens due to spontaneous condensation of bound electron-hole pairs. Ta$_2$NiSe$_5$ is a promising candidate for this type of material, but the coexistence of a structural phase transition with the gap opening has led to a long-standing debate regarding the origin of the insulating gap. Here we employ MeV ultrafast electron diffraction to obtain quantitative insights into the atomic displacements in Ta$_2$NiSe$_5$ following photoexcitation, which has been overlooked in previous time-resolved spectroscopy studies. In conjunction with first-principles calculations using the measured atomic displacements,  we find that the structural change can largely account for the photoinduced reduction in the energy gap without considering excitonic effects. Our work illustrates the importance of a quantitative reconstruction of individual atomic pathways during nonequilibrium phase transitions, paving the way for a mechanistic understanding of a diverse array of phase transitions in correlated materials where lattice dynamics can play a pivotal role.
\end{abstract}

\maketitle

\begin{figure*}[t!]
    \includegraphics[width=1\textwidth]{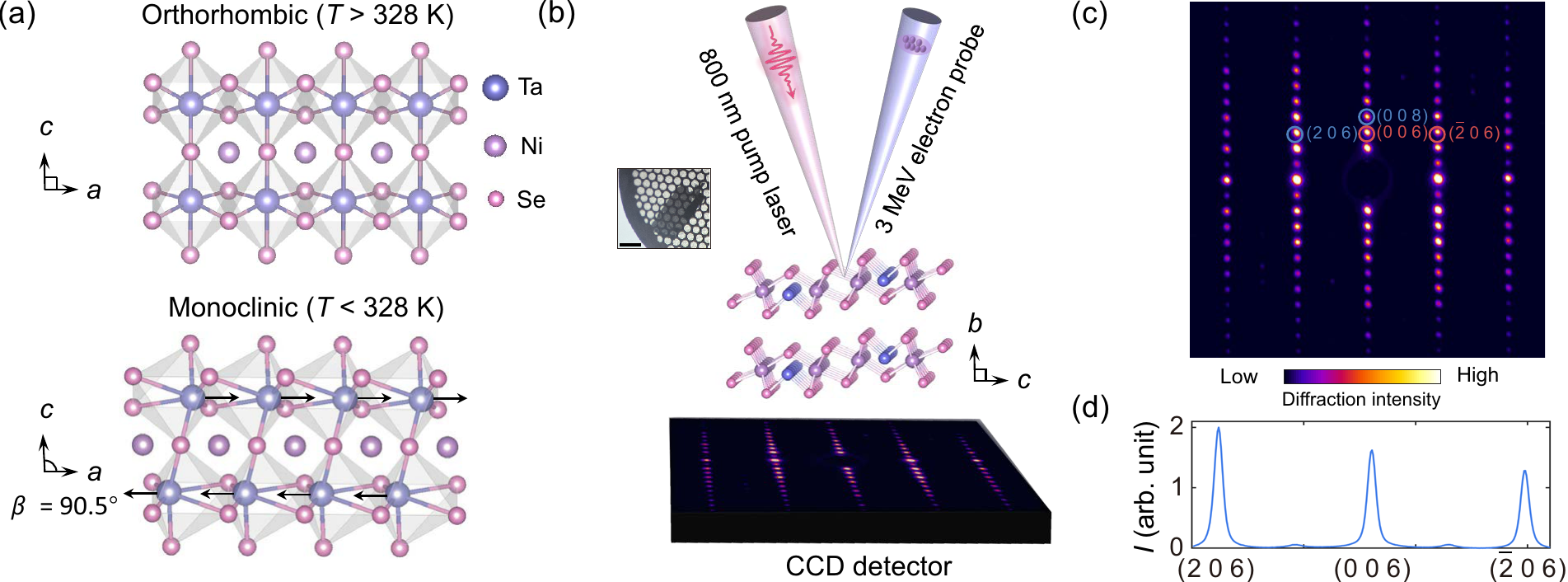}
    \caption{(a)~Crystal structure of Ta$_2$NiSe$_5$ in the high-temperature orthorhombic and low-temperature monoclinic phase. The arrows in the bottom panel indicate the relative displacement of Ta and Se atoms along the $a$-axis, where the arrow length is exaggerated for visual clarity. (b)~Schematic of the MeV-UED setup.~\textit{Inset}: Optical image of the sample supported on copper grids with an ultrathin carbon film ($<$~10~nm); scale bar:~200~$\upmu\mathrm{m}$. See Sec.~I of the Supplemental Material ~\cite{*[{See Supplemental Material at }] [{for more information on these topics, which contains Refs. [21–32]}] SupplementalMaterial} for sample preparation details. (c)~Static electron diffraction pattern taken at  200~K with the 3~MeV electrons.  (d)~Intensity profiles of three representative diffraction peaks along the $(h06)$ cut. }	
    \label{Fig1}
\end{figure*}

The phase of an excitonic insulator (EI), whereby bound electron-hole pairs spontaneously condense to open an energy gap in the ground state, has been predicted to occur in semiconductors with a small band gap or semimetals with a small band overlap~\cite{mott1961transition,kohn1967excitonic,halperin1968possible}. To date, only a limited number of bulk crystalline materials have been proposed as potential hosts for a ground state EI phase, including TiSe$_2$~\cite{cercellier2007evidence,monney2011exciton,kogar2017signatures},  Ta$_2$NiSe$_5$~\cite{wakisaka2009excitonic,kaneko2013orthorhombic,seki2014excitonic}, TmSe$_{0.45}$Te$_{0.55}$~\cite{bucher1991excitonic}, and Ta$_2$Pd$_3$Te$_5$~\cite{Zhang2024spontaneous,huang2024evidence,hossain2025topological}. Of particular interest is the material Ta$_2$NiSe$_5$, which exhibits a small direct band gap, no instability at nonzero wave vector, and a high transition temperature. Upon cooling to a critical temperature of $T_{\textrm
{c}}$~=~328~K, an energy gap opens in the electronic band structure with an anomalous flattening and broadening of the valance band~\cite{wakisaka2009excitonic}, consistent with the scenario of formation of EI phase. However, the band gap opening in \tns~is accompanied by a simultaneous structural phase transition~\cite{sunshine1985structure,di1986physical,kim2016layer,chen2023role} (as shown in Fig.~\ref{Fig1}(a)),
which can influence the gap through structural distortion-induced hybridization of the conduction and valence bands ~\cite{nakano2018antiferroelectric,mazza2020nature,chen2023role}.
The interplay of the changing energy gap and structural distortion as a function of temperature makes it challenging to ascertain the true nature of gap opening through measurements under equilibrium conditions. For instance, clear signatures of excitonic instability across  $T_c$ were revealed by temperature-dependent Raman spectroscopy,highlighting the significant role of excitonic correlations in \tns~\cite{kim2021direct,volkov2021critical}. By contrast, angle-resolved photoemission spectroscopy (ARPES) measurements have proposed that lattice instability is the primary driving force of the gap opening, with excitonic correlations playing a secondary role~\cite{watson2020band,chen2023role}.

Ultrafast pump-probe methods offer significant potential for elucidating the intricate interplay between different degrees of freedom, as they allow the separate measurements of electronic and structural dynamics that may occur on different timescales or respond differently to photoexcitation~\cite{hellmann2012time}. To this end, numerous ultrafast spectroscopic measurements have been employed to investigate the photoinduced dynamics of \tns~\cite{mor2017ultrafast,okazaki2018photo,tang2020non,suzuki2021detecting,saha2021photoinduced,baldini2023spontaneous,bretscher2021ultrafast,miyamoto2022charge,katsumi2023disentangling,werdehausen2018coherent,Mor2022ultrafast,miyamoto2022charge}, but conflicting results were obtained in regard to whether the transition is excitonic or structural in nature. For instance, many studies using time-resolved ARPES (tr-ARPES)~\cite{tang2020non,saha2021photoinduced,suzuki2021detecting,golevz2022unveiling} have observed a reduction and possible closure of the energy gap upon photoexcitation, a process that is accompanied by the presence of a 2~THz coherent phonon. This phonon mode, as identified in previous Raman and optical spectroscopy measurements under thermal equilibrium~\cite{werdehausen2018coherent,kim2021direct,volkov2021critical,bretscher2021ultrafast}, is well-defined only in the low-temperature monoclinic phase and becomes highly damped in the high-temperature phase. The concurrent observation of this 2~THz phonon and gap closure is interpreted as evidence of a transient semimetallic state without a structural phase transition, supporting an excitonic origin for the band gap opening in \tns. By contrast, a recent study combined tr-ARPES measurements with first-principles calculations and proposed a structural origin for the gap~\cite{baldini2023spontaneous}. This experiment was conducted under a condition where the electronic temperature exceeded $T_c$, while the lattice temperature remained below $T_c$. The persistence of the energy gap under these conditions lends support to the notion that a structural mechanism is responsible for driving the phase transition.

The inconsistent results and interpretation from ultrafast spectroscopic probes stem from a lack of direct access to the structural degree of freedom. In these experiments, the role of phonons and structural distortions can only be indirectly inferred from their impact on the electronic dispersion, the timescale of the photoinduced dynamics, and the coherent modes excited by the laser pulse. To reach a definitive conclusion about the nature of the photoinduced transition, it is therefore crucial to quantitatively determine the atomic trajectories. To this end, we use MeV-ultrafast electron diffraction (MeV-UED) to investigate the structural dynamics of \tns. The 50-fs temporal resolution of MeV-UED set up~\cite{qi2020breaking} enables us to resolve the fastest photoinduced changes that occur over 200~fs in \tns~\cite{okazaki2018photo,tang2020non,suzuki2021detecting,saha2021photoinduced,bretscher2021ultrafast,miyamoto2022charge}. Furthermore, the large momentum space accessible with MeV electrons allows us to detect nearly 100 Bragg peaks, which makes it possible to quantitatively map out motions of each type of atoms with high accuracy. We observed that the low-temperature monoclinic structural distortion was reduced upon photoexcitation. First-principles calculation using the photoinduced atomic displacements shows a clear reduction in the band gap, which is in good agreement with that measured in tr-ARPES. {Our ability to reproduce most of the electronic band evolutions observed in tr-ARPES by calculating the spectral gap from the measured atomic displacements --- without incorporating excitonic effects --- suggests that the photoinduced gap dynamics in Ta$_2$NiSe$_5$ is primarily related to structural changes.} 

In \tns, the low temperature monoclinic structural distortion manifests as a shear of the Ta and Se atoms along the $a$ direction, as indicated by the black arrows in Fig.~\ref{Fig1}(a). This shear distortion is related to the $A_{g}$ zone-center optical phonon with a frequency of approximately 2~THz, which is widely considered to be the signature of the monoclinic phase~\cite{subedi2020orthorhombic}. The schematic of the pump-probe experiments is illustrated in Fig.~\ref{Fig1}(b). The dynamics are initiated with above-gap photoexcitation using an 800~nm laser pulse, which is similar to the previous tr-ARPES experiments~\cite{tang2020non,baldini2023spontaneous,saha2021photoinduced,suzuki2021detecting}, and the structural dynamics are probed with a high-energy electron pulse. {The penetration depth of the 800-nm pump is approximately 55~nm~\cite{liu2021photoinduced}, comparable to the sample thickness (50~nm).} The representative static electron diffraction pattern collected at 200~K is shown in Fig.~\ref{Fig1}(c). The sharp Bragg peaks indicate the high quality of the sample. As shown in Fig.~\ref{Fig1}(d), the (206) peak exhibits a higher intensity than the ($\overline{2}$06) peak at 200~K, which is a direct consequence of the monoclinic distortion. Without monoclinic distortion above $T_c$, the two peaks are expected to have equal intensity (see Fig.~S1 in the Supplemental Material~\cite{*[{See Supplemental Material at }] [{for more information on these topics, which contains Refs. [21–32].}] SupplementalMaterial}).

\begin{figure} [t!] 
    \includegraphics
[width=\columnwidth]{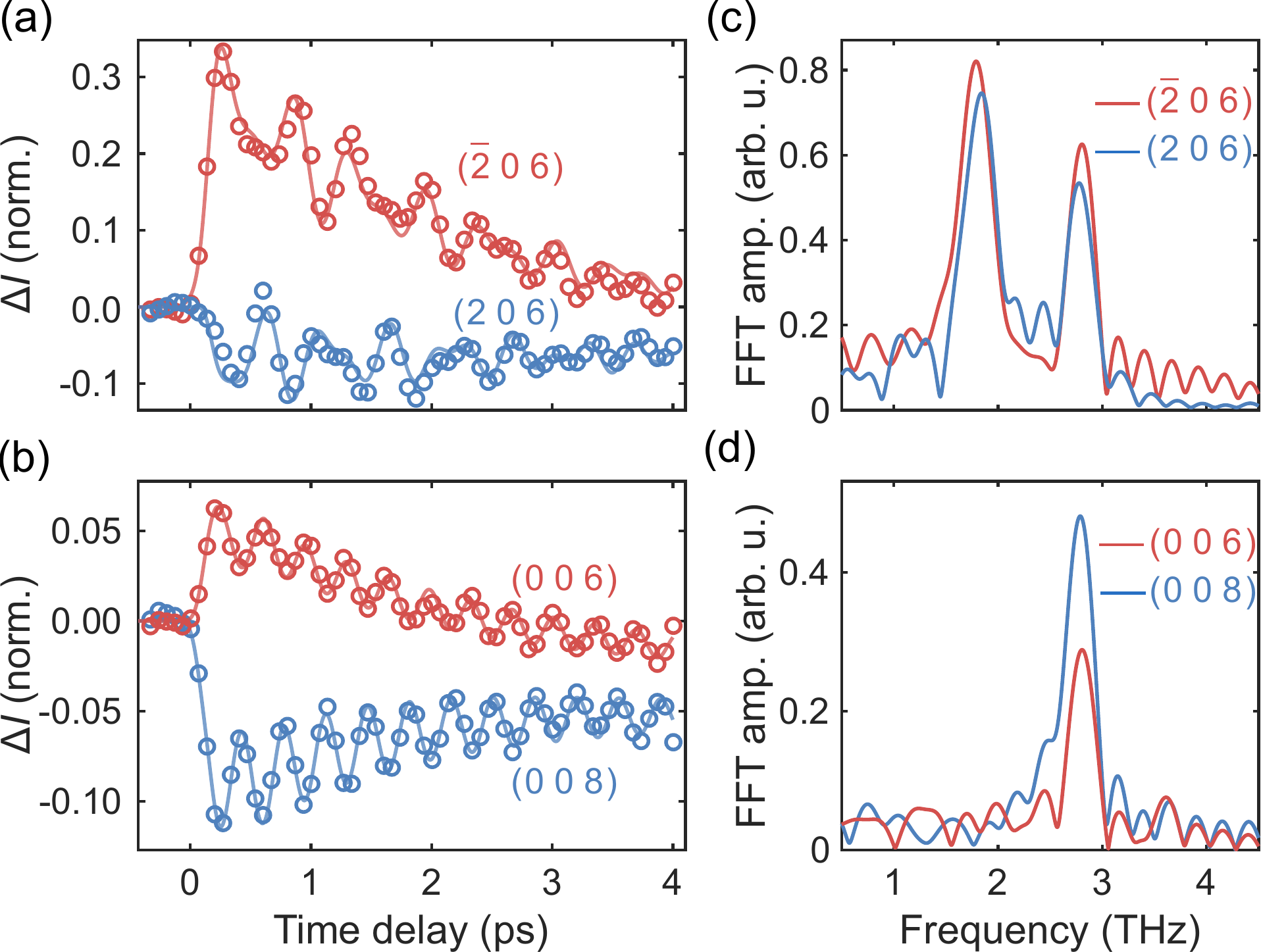}
    \caption{ (a,b)~Temporal evolution of peak intensity  for selected Bragg orders ($\overline{2}$06), (206), (006) and (008) for the pump fluence of 1.6~\mjc. The change is normalized to the averaged values prior to photoexcitation of the respective peaks. The solid curves are fits to Eq.~(1) in Supplemental Material~\cite{*[{See Supplemental Material at }] [{for more information on these topics, which contains Refs. [21–32].}] SupplementalMaterial}. (c,d)~The corresponding Fourier transform magnitudes of the oscillatory components in (a) and (b), where the time traces were zero-padded before the Fast Fourier Transform (FFT) operation.}	
    \label{Fig2}
\end{figure}
 
The photoinduced intensity change of a few representative Bragg peaks is shown in Fig.~\ref{Fig2}(a)(b), where the incident fluence was 1.6~mJ/cm$^2$. {Upon photoexcitation, the (206) peak undergoes a decrease in intensity, while the ($\overline{2}$06) peak experiences an increase, a trend that is independent from the pump laser polarization (see Fig.~S5 in the Supplemental Material~\cite{*[{See Supplemental Material at }] [{for more information on these topics, which contains Refs. [21–32].}] SupplementalMaterial}).  These intensity modulations are attributable to the Ta and Se atoms moving towards the high-temperature orthorhombic phase along the $a$-axis, which results in the reduction of monoclinic distortion. Similarly, the (006) and (008) peaks exhibit opposite behaviors upon photoexcitation. While the intensity of the (006) peak increases, the intensity of the (008) peak decreases. This phenomenon can be attributed to the photoinduced 3~THz coherent phonon, which primarily involves the oscillatory motion of Ta and Se atoms along the $c$-axis~\cite{guan2023coherent}. In order to make a qualitative comparison, the electron diffraction intensity modulations induced by the two kinds of atomic displacements were simulated [see Fig.~S3(a) and (b)~\cite{*[{See Supplemental Material at }] [{ for extensive derivation of theory and simulation and experimental
details.}] SupplementalMaterial}], and the simulation results were in close agreement with our observations.
   
To extract the frequencies of these oscillations, we plot the amplitude of the Fast Fourier Transform~(FFT) of these peak intensities in Fig.~\ref{Fig2}(c)(d). The (206) and ($\overline{2}$06) peaks exhibit clear oscillations at approximately 2~THz and 3~THz. These two phonon modes have been observed by tr-ARPES and ultrafast optical spectroscopy measurements~\cite{werdehausen2018coherent,tang2020non,golevz2020nonlinear,bretscher2021ultrafast,suzuki2021detecting,kim2021direct}. {However, two other phonons at 1 THz and 4 THz, which were reported in ultrafast spectroscopy, are absent in our diffraction data. For the 1~THz phonon, its absence is due to the fact that the incident electrons are normal to the $ac$-plane of \tns, whereas the 1~THz phonon is mainly associated with atomic motion along the $b$-axis~\cite{werdehausen2018coherent}. For the 4~THz phonon, based on structure factor calculations, the projected atomic motions in the $ac$ plane do not substantially modulate the diffraction intensities of the Bragg peaks of interest, so it remains undetectable given the signal-to-noise ratio of the present experiment. In contrast to the (206) and ($\overline{2}$06) peaks}, a single frequency oscillation at approximately 3~THz is observed for the (006) and (008) peaks, as illustrated in Figs.~\ref{Fig2}(b) and \ref{Fig2}(d). The calculated eigenvectors \cite{suzuki2021detecting} of the dynamical matrix of \tns~indicate that the atomic motions for the 2~THz phonon mainly exhibit a shear along the $a$-axis, while those for the 3~THz phonon mainly oscillate along the $c$-axis~\cite{guan2023coherent}. Consequently, the 2~THz phonon only modulates the structure factors of the $(hkl)$ peaks with a non-zero $h$ value such as ($\pm206$), whereas the 3~THz phonon only modulates the structure factors of the $(hkl)$ peaks with a non-zero $l$ value such as (006) and (008). Here, $(hkl)$ represents the Miller indices of the diffraction peaks.

To gain further insights into the photoinduced structural dynamics, we employed a global fitting approach to quantitatively determine the ultrafast atomic motions following photoexcitation (see the Supplemental Material for analysis details~\cite{*[{See Supplemental Material at }] [{for more information on these topics, which contains Refs. [21–32].}] SupplementalMaterial}). The measured diffraction intensity change at $t$~=~0.2~ps is shown in Fig.~\ref{Fig3}(a), where the photoinduced change is maximal. The corresponding simulated diffraction intensity change with the best-fit atomic positions obtained through global fitting is shown in Fig.~\ref{Fig3}(b). For a quantitative comparison, the measured and simulated diffraction intensities for the most intense peaks are presented in Fig.~\ref{Fig3}(c), showing excellent agreement.
 
\begin{figure}[t!]
    \includegraphics[width=\columnwidth]{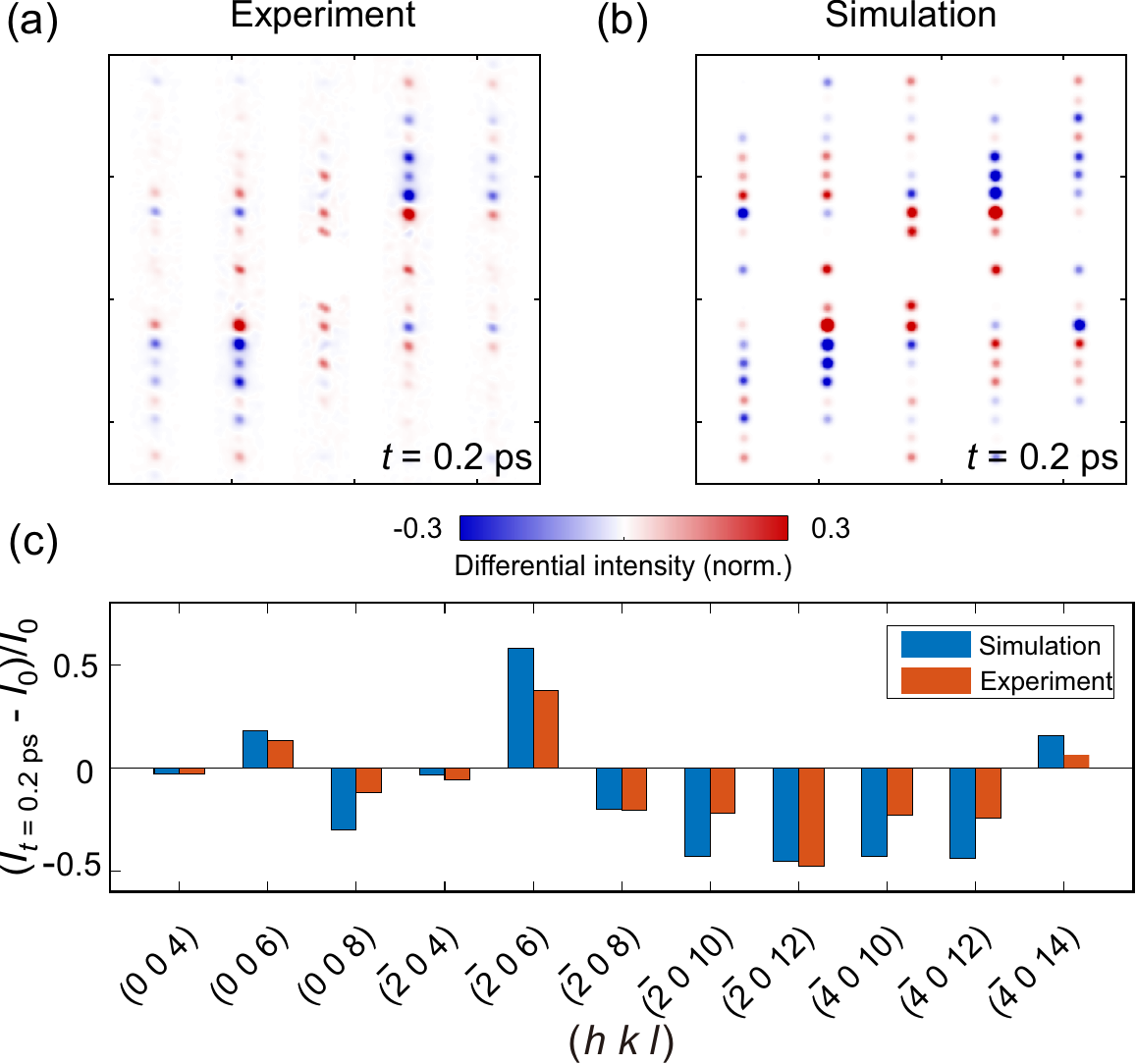}
    \caption{(a)~Measured electron diffraction intensity changes at 0.2~ps.  (b)~Simulated diffraction intensity changes by displacing the Ta and Se atoms according to the atomic motion of the 2 and 3~THz phonons, as determined from global fitting. (c)~Comparison of measured and simulated diffraction intensity changes at 0.2~ps for representative Bragg peaks.  }
    \label{Fig3}
\end{figure}

We next applied the global-fitting method to determine the corresponding atomic displacements at other time delays, which can offer an atomic view of the complete photoinduced structural dynamics. In the main text, we focus on the displacement of the Ta atoms, and the information on other atoms can be found in the Supplemental Material~\cite{*[{See Supplemental Material at }] [{for more information on these topics, which contains Refs. [21–32].}] SupplementalMaterial}. The changes in the displacement of Ta atoms along the \( c \)-axis~(\( \Delta\text{Ta}_c \)) and \( a \)-axis~(\( \Delta\text{Ta}_a \)) as a function of time delay for three different pump fluences are illustrated in Figs.~4(a) and (b), respectively. It can be seen that the change in displacement reaches a maximal value at approximately $t = 0.2~$ps and then recovers on a timescale of a few picoseconds. During the recovery phase, the Ta atom exhibits oscillatory behavior at approximately 2~THz and 3~THz along the $a$-axis and $c$-axis, respectively. This motion is consistent with the calculated phonon eigenvectors
~\cite{suzuki2021detecting}, which is challenging to be revealed by spectroscopic methods. 

In light of the photoinduced metastable state of Ta$_2$NiSe$_5$ presented in ref.~\cite{liu2021photoinduced}, we have incorporated the suggested shear motion of each layer in the $ab$ plane into our global fitting approach. This motion has been previously observed to result in a permanent phase transition in Ta$_2$NiSe$_5$ when the pump fluence is sufficiently high. It is important to note that the displacement associated with the shear motion is significantly smaller in our experimental observations compared to that reported in previous studies (see the Supplemental Material for details~\cite{*[{See Supplemental Material at }] [{ for more information on these topics, which contains Refs. [21–32].}] SupplementalMaterial}). This observation suggests that our pump-probe experiment remains within the fully reversible regime.

\begin{figure}[b!]
\includegraphics[width=\columnwidth]{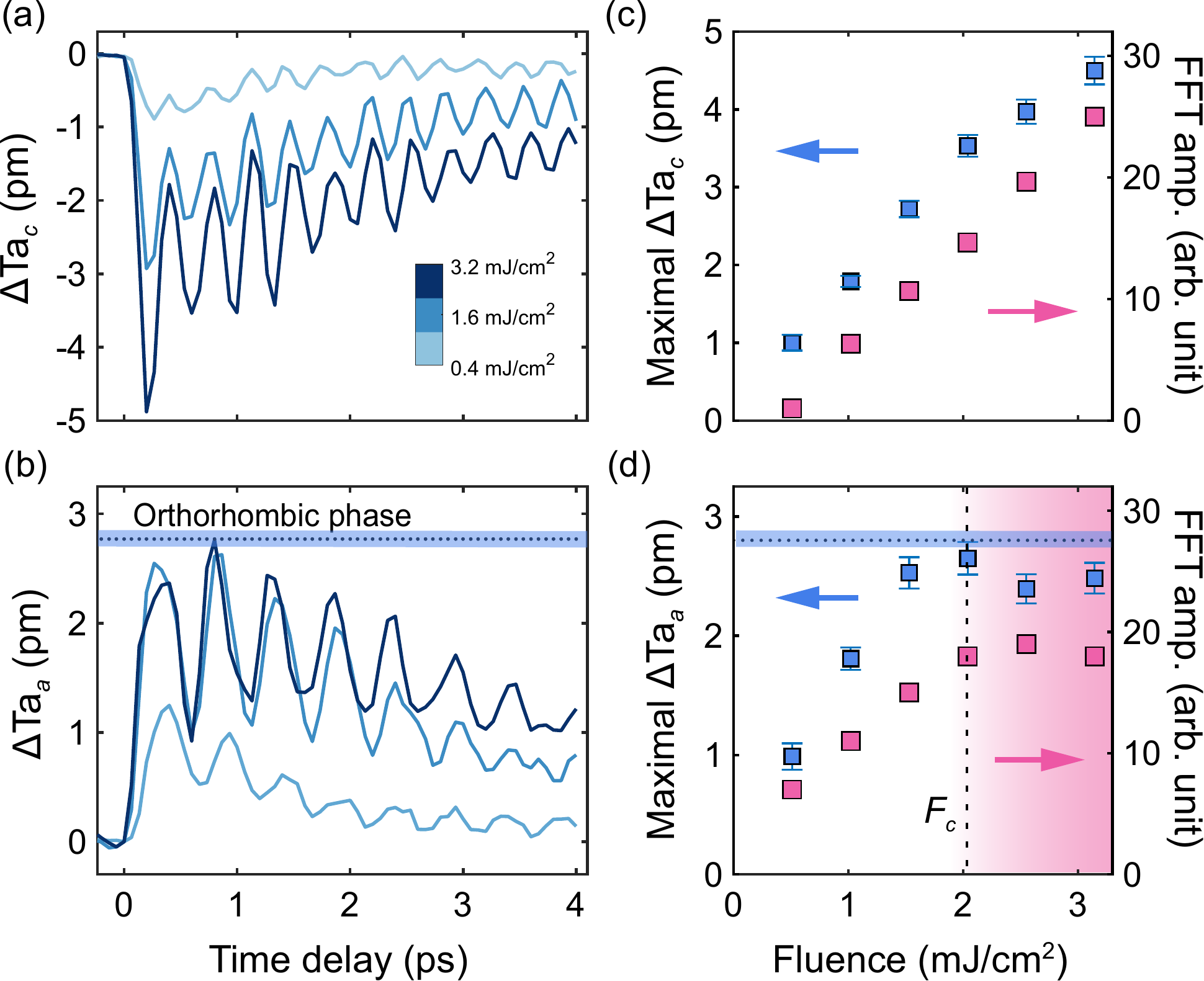}
    \caption{(a,b)~Retrieved time-dependent displacement of the Ta atom along the $c$-axis (a) and $a$-axis (b) at three pump fluences, respectively. (c,d)~The pump fluence dependence of the coherent phonon amplitude (taken as the FFT amplitude) and the maximal Ta displacement along the $c$-axis (c) and $a$-axis (d), respectively. In (b,d), the horizontal dashed line indicates the equilibrium Ta displacement of $2.77\pm0.05$~pm along the $a$-axis due to the structural transition \cite{sunshine1985structure}, where the line thickness denotes the uncertainty.}
    \label{Fig4}
\end{figure}

Despite the concurrent excitation of both the 2~THz and 3~THz phonons, the lattice displacement along $a$- and $c$-axis exhibit distinct fluence-dependent behaviors. As illustrated in Fig.~\ref{Fig4}(a), the displacement along the $c$-axis exhibits a pronounced increase when the pump fluence is increased from 1.6 to 3.2 mJ/cm$^2$. By contrast, the displacement along $a$-axis [Fig.~\ref{Fig4}(b)] shows a negligible difference between these two fluences. To illustrate the distinction more clearly, we present the maximum displacements obtained with six different pump fluences in Figs.~4(c)~and~4(d), which are overlaid with the oscillation amplitudes from FFT. As illustrated in Figs.~4(b) and 4(d), upon photoexcitation, the Ta atoms move towards the high-temperature phase, thereby reducing the monoclinic distortion. However, as the pump fluence increases, the maximal displacement of the Ta atoms along the $a$-axis reaches a saturation value at about  $\Delta\text{Ta}_a = 2.6$~pm. This value is comparable to the displacement calculated from the static measurement taken during a complete phase transition at thermal equilibrium ($2.77\pm0.05$~pm~\cite{sunshine1985structure}), indicated by the horizontal dotted lines in Fig.~\ref{Fig4}(b) and \ref{Fig4}(d). The saturation of $\Delta\text{Ta}_a$ echoes the saturation of the oscillation amplitude of the 2~THz phonon (Fig.~\ref{Fig4}(d)), which has been observed in a number of ultrafast spectroscopic measurements~\cite{mor2017ultrafast, Mor2018inhibition, Werdehausen2018photo, tang2020non, bretscher2021ultrafast, werdehausen2018coherent}. The phonon saturation was theoretically attributed to the saturation of the photoexcited carriers from the flat valance band top~\cite{mor2017ultrafast}, which results in the atomic displacement and amplitude saturation of 2~THz phonon mode due to exciton-phonon coupling~\cite{guan2023coherent, ning2020signatures}. In this scenario, the exciton exhibits a strong coupling with the 2~THz phonon, such that once the exciton is completely depleted by the pump photons, both the atomic displacement and the oscillation amplitude reach saturation values. In light of our quantitative determination of the atomic displacement through MeV-UED, at a fluence value beyond 1.6~mJ/cm$^2$, the Ta atoms have already returned to their high-temperature position along the $a$-axis, therefore exhibiting a saturation behavior.

On the other hand, as shown in Fig.~\ref{Fig4}(a) and \ref{Fig4}(c), the displacements along the $c$-axis and the amplitude of the 3~THz phonon increase in a linear fashion as the laser fluence rises. This phonon is observed to persist in both the low-temperature monoclinic phase and the high-temperature orthorhombic phase, and is not coupled to the exciton~\cite{guan2023coherent}. Consequently, both the atomic displacement and the oscillation amplitude increase in line with the production of additional carriers through alternative transition pathways, as the pump fluence is increased. The lack of fluence saturation in $\Delta\text{Ta}_c$ is indicative of the lack of structural constraints along the $c$-direction for the monoclinic-to-orthorhombic transition.

\begin{figure} [b!]    
    \centering    \includegraphics[width=\columnwidth]{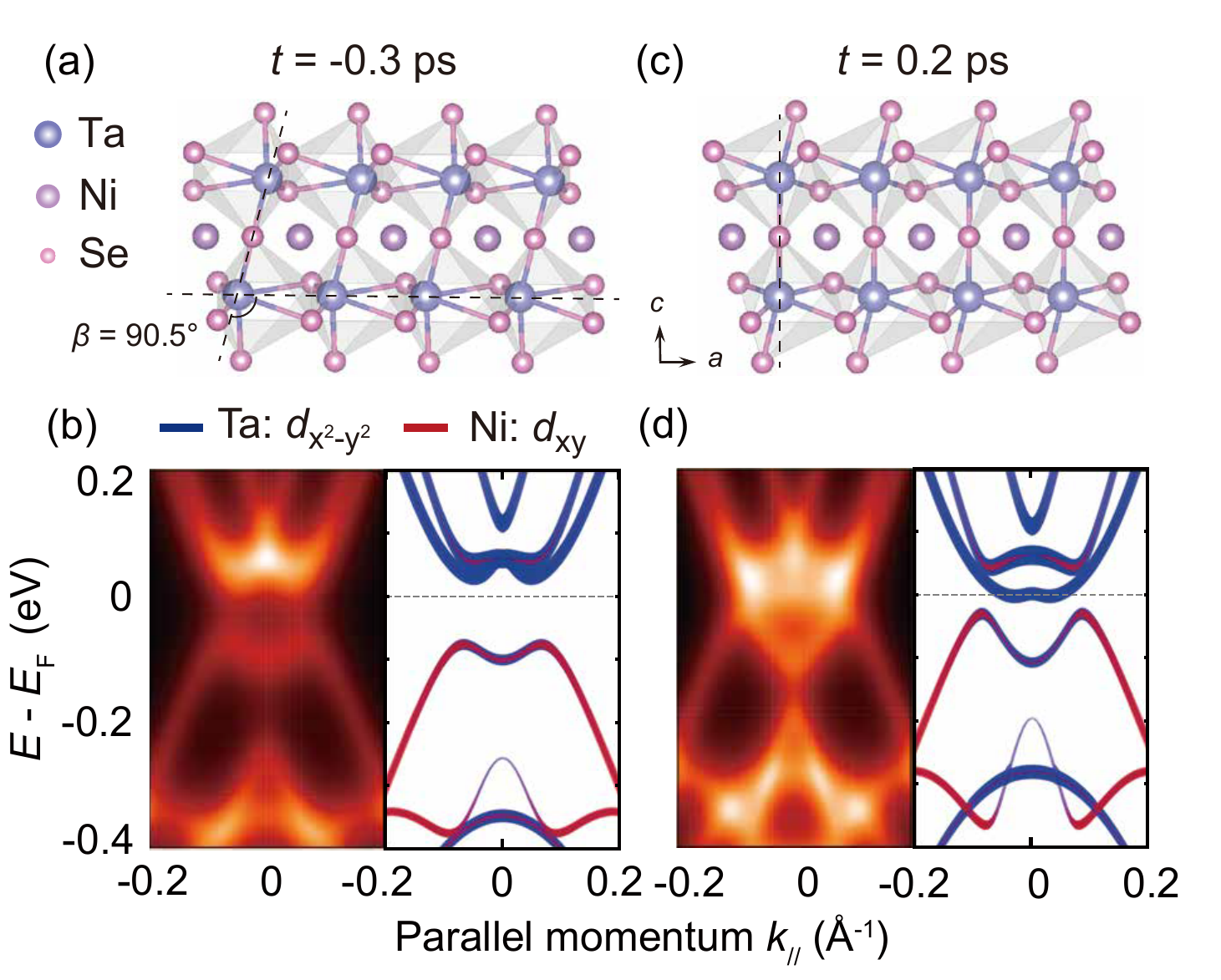}
    \caption{The simulated crystal structure (with exaggerated displacements), single-particle spectra with Cauchy-Lorentzian convolution, and orbital-resolved non-equilibrium band structure along the X-$\Gamma$-X direction of Ta$_2$NiSe$_5$ by DFT calculation using a Fermi-Dirac smearing at $t = -0.3~$ps (a--b) and $t = 0.2~$ps (c--d), respectively. Ta $d_{\mathrm{x^2-y^2}}$ (blue) and Ni $d_{\mathrm{xy}}$ (red) orbitals are highlighted. Line thickness corresponds to the spectral weight.}	
    \label{Fig5}
\end{figure}  

{ Atomic displacements are expected to induce renormalization of the electronic structure in \tns. Based on the maximally displaced transient lattice structure deduced from the electron diffraction pattern at 0.2~ps under the pump fluence of 1.6~mJ/cm$^2$, we computed the transient single-particle spectra and orbital-resolved band structure using density-functional theory (DFT), and contrasted them to those prior to photoexcitation, shown in Fig.~\ref{Fig5} (see Sec.~XII in the Supplemental Material for computation details~\cite{*[{See Supplemental Material at }] [{ for more information on these topics, which contains Refs. [21–32].}] SupplementalMaterial}). In these calculations, we made the quasi-equilibrium approximation. Specifically, we used the deformation potential obtained from the equilibrium state to calculate the electronic structure based on the atomic displacements obtained from our measurements, which stay in the linear response regime below the threshold fluence. A similar assumption was made in the context of coherently excited phonons to extract electron-phonon couplings \cite{gerber2017femtosecond,huang2023ultrafast}, which yield good agreements between experimental observations and first-principles calculations.}


As illustrated in Fig.~\ref{Fig5}(a), the system begins in a monoclinic state with a lattice order parameter $\beta$ comparable [see schematic in Fig.~\ref{Fig1}(a)] with a gap of $\sim100$~meV around $E_F$, consistent with previous reports~\cite{chen2023role,chen2023anomalous}. At 0.2~ps, the measured transient lattice structure displayed in Fig.~\ref{Fig5}(c) yields an electronic dispersion where the gap has been substantially reduced, and the flat band top regains its curvature characteristic of the high-temperature phase. This change indicates that the structure transitions to a near-orthorhombic semimetallic phase, consistent with prior tr-ARPES measurements~\cite{tang2020non,saha2021photoinduced,suzuki2021detecting}. Importantly, the orbital composition of the 0.2~ps electronic structure in Figs.~\ref{Fig5}(b) and \ref{Fig5}(d) is also in excellent agreement with the measured bands in earlier static photoemission experiments conducted above $T_c$~\cite{chen2023role}, lending further evidence for a transient semimetallic state driven by the lattice distortion. 

Previous frequency-domain tr-ARPES studies have indicated that the valence band top is mainly modified by the 2~THz phonon~\cite{suzuki2021detecting}. Another tr-ARPES study on Ta$_2$NiSe$_5$ reported an abrupt shift in the valence band towards the Fermi level upon photoexcitation,  followed by few-picosecond relaxation and pronounced 2~THz oscillations ~\cite{tang2020non}. The aforementioned tr-ARPES measurements, which adopt an electronic perspective, are aligned with our structural calculations based on atomic motions extracted from UED.  Notably, the valence band shift reaches its maximum energy at 0.1--0.3~ps, which agrees with the timescale of initial atomic motions in Fig.~\ref{Fig4}(a,b). Furthermore, the narrowing of the energy gap also displays saturation behavior, albeit with a lower saturation threshold \cite{mor2017ultrafast,tang2020non}. {This saturation behavior has also been observed in other spectroscopic measurements, where varying fluence thresholds were reported \cite{Werdehausen2018photo,werdehausen2018coherent,katsumi2023disentangling}.}; it is not uncommon for fluence value discrepancies to arise in tr-ARPES and diffraction studies due to the differing experimental conditions (see, for example, \cite{gerber2017femtosecond,huang2023ultrafast}). Importantly, these previously reported nonequilibrium dynamics exhibit a striking resemblance to the evolution of \( \Delta\text{Ta}_a \) obtained from UED and the calculated band gap evolution. In light of this consistency, we therefore conclude that the evolution of the band gap is mainly driven by the lattice structure transition, which manifests as the displacements of Ta and Se atoms along the $a$-axis (see Sec.~X in the Supplemental Material for an analysis of the Se atoms~\cite{*[{See Supplemental Material at }] [{ for more information on these topics, which contains Refs. [21–32].}] SupplementalMaterial}). It should be noted that we cannot entirely rule out the possibility of exciton condensation and its effect on the photoinduced dynamics because not all observations from time-resolved optical spectroscopy and ARPES studies are fully reconciled by our experiments \cite{mor2017ultrafast,golevz2022unveiling,saha2021photoinduced,okazaki2018photo}. However, given that our structural explanation accounts for the key observations of gap suppression and the associated fluence saturation, we conclude that structural dynamics play the primary role.

To conclude, by leveraging MeV-UED to access a large number of diffraction peaks with superior temporal resolution, we offer a new, structurally-based perspective on the photoinduced dynamics in \tns. We observed that the transient structural distortion that resembles the orthorhombic state can largely account for the observed band structure change in previous time-resolved photoemission experiments, as verified by our first-principles calculations using the quantitatively determined atomic trajectories. It is shown that photoexcitation can partially restore both the electronic and lattice structure from a broken-symmetry monoclinic state to a nearly-symmetric orthorhombic state, suggesting that the structural transformation cannot be overlooked in the energy gap dynamics of Ta$_2$NiSe$_5$.

\begin{acknowledgments}
This work is supported by the National Natural Science Foundation of China (Nos.~12525501, 12335010 and 12488201) and the National Key R$\&$D Program of China (No.~2021YFA1400202, 2022YFA1403901, 2022YFA1402400 and 2024YFA1612204). D.X. would like to acknowledge the support from the New Cornerstone Science Foundation through the Xplorer Prize. The UED experiment was supported by the Shanghai soft X-ray free-electron laser facility. This work was also supported by the Synergetic Extreme Condition User Facility (SECUF).
\end{acknowledgments}


\begin{thebibliography}{47}%
\makeatletter
\providecommand \@ifxundefined [1]{%
 \@ifx{#1\undefined}
}%
\providecommand \@ifnum [1]{%
 \ifnum #1\expandafter \@firstoftwo
 \else \expandafter \@secondoftwo
 \fi
}%
\providecommand \@ifx [1]{%
 \ifx #1\expandafter \@firstoftwo
 \else \expandafter \@secondoftwo
 \fi
}%
\providecommand \natexlab [1]{#1}%
\providecommand \enquote  [1]{``#1''}%
\providecommand \bibnamefont  [1]{#1}%
\providecommand \bibfnamefont [1]{#1}%
\providecommand \citenamefont [1]{#1}%
\providecommand \href@noop [0]{\@secondoftwo}%
\providecommand \href [0]{\begingroup \@sanitize@url \@href}%
\providecommand \@href[1]{\@@startlink{#1}\@@href}%
\providecommand \@@href[1]{\endgroup#1\@@endlink}%
\providecommand \@sanitize@url [0]{\catcode `\\12\catcode `\$12\catcode `\&12\catcode `\#12\catcode `\^12\catcode `\_12\catcode `\%12\relax}%
\providecommand \@@startlink[1]{}%
\providecommand \@@endlink[0]{}%
\providecommand \url  [0]{\begingroup\@sanitize@url \@url }%
\providecommand \@url [1]{\endgroup\@href {#1}{\urlprefix }}%
\providecommand \urlprefix  [0]{URL }%
\providecommand \Eprint [0]{\href }%
\providecommand \doibase [0]{https://doi.org/}%
\providecommand \selectlanguage [0]{\@gobble}%
\providecommand \bibinfo  [0]{\@secondoftwo}%
\providecommand \bibfield  [0]{\@secondoftwo}%
\providecommand \translation [1]{[#1]}%
\providecommand \BibitemOpen [0]{}%
\providecommand \bibitemStop [0]{}%
\providecommand \bibitemNoStop [0]{.\EOS\space}%
\providecommand \EOS [0]{\spacefactor3000\relax}%
\providecommand \BibitemShut  [1]{\csname bibitem#1\endcsname}%
\let\auto@bib@innerbib\@empty
\bibitem [{\citenamefont {{Mom}}(1961)}]{mott1961transition}%
  \BibitemOpen
  \bibfield  {author} {\bibinfo {author} {\bibfnamefont {N.~F.}\ \bibnamefont {{Mom}}},\ }\bibfield  {title} {\bibinfo {title} {{The transition to the metallic state}},\ }\href {https://doi.org/10.1080/14786436108243318} {\bibfield  {journal} {\bibinfo  {journal} {Philosophical Magazine}\ }\textbf {\bibinfo {volume} {6}},\ \bibinfo {pages} {287} (\bibinfo {year} {1961})}\BibitemShut {NoStop}%
\bibitem [{\citenamefont {{Kohn}}(1967)}]{kohn1967excitonic}%
  \BibitemOpen
  \bibfield  {author} {\bibinfo {author} {\bibfnamefont {W.}~\bibnamefont {{Kohn}}},\ }\bibfield  {title} {\bibinfo {title} {{Excitonic Phases}},\ }\href {https://doi.org/10.1103/PhysRevLett.19.439} {\bibfield  {journal} {\bibinfo  {journal} {\prl}\ }\textbf {\bibinfo {volume} {19}},\ \bibinfo {pages} {439} (\bibinfo {year} {1967})}\BibitemShut {NoStop}%
\bibitem [{\citenamefont {{Halperin}}\ and\ \citenamefont {{Rice}}(1968)}]{halperin1968possible}%
  \BibitemOpen
  \bibfield  {author} {\bibinfo {author} {\bibfnamefont {B.~I.}\ \bibnamefont {{Halperin}}}\ and\ \bibinfo {author} {\bibfnamefont {T.~M.}\ \bibnamefont {{Rice}}},\ }\bibfield  {title} {\bibinfo {title} {{Possible Anomalies at a Semimetal-Semiconductor Transistion}},\ }\href {https://doi.org/10.1103/RevModPhys.40.755} {\bibfield  {journal} {\bibinfo  {journal} {Rev. Mod. Phys.}\ }\textbf {\bibinfo {volume} {40}},\ \bibinfo {pages} {755} (\bibinfo {year} {1968})}\BibitemShut {NoStop}%
\bibitem [{\citenamefont {{Cercellier}}\ \emph {et~al.}(2007)\citenamefont {{Cercellier}}, \citenamefont {{Monney}}, \citenamefont {{Clerc}}, \citenamefont {{Battaglia}}, \citenamefont {{Despont}}, \citenamefont {{Garnier}}, \citenamefont {{Beck}}, \citenamefont {{Aebi}}, \citenamefont {{Patthey}}, \citenamefont {{Berger}},\ and\ \citenamefont {{Forr{\'o}}}}]{cercellier2007evidence}%
  \BibitemOpen
  \bibfield  {author} {\bibinfo {author} {\bibfnamefont {H.}~\bibnamefont {{Cercellier}}}, \bibinfo {author} {\bibfnamefont {C.}~\bibnamefont {{Monney}}}, \bibinfo {author} {\bibfnamefont {F.}~\bibnamefont {{Clerc}}}, \bibinfo {author} {\bibfnamefont {C.}~\bibnamefont {{Battaglia}}}, \bibinfo {author} {\bibfnamefont {L.}~\bibnamefont {{Despont}}}, \bibinfo {author} {\bibfnamefont {M.~G.}\ \bibnamefont {{Garnier}}}, \bibinfo {author} {\bibfnamefont {H.}~\bibnamefont {{Beck}}}, \bibinfo {author} {\bibfnamefont {P.}~\bibnamefont {{Aebi}}}, \bibinfo {author} {\bibfnamefont {L.}~\bibnamefont {{Patthey}}}, \bibinfo {author} {\bibfnamefont {H.}~\bibnamefont {{Berger}}},\ and\ \bibinfo {author} {\bibfnamefont {L.}~\bibnamefont {{Forr{\'o}}}},\ }\bibfield  {title} {\bibinfo {title} {{Evidence for an Excitonic Insulator Phase in 1T-TiSe$_{2}$}},\ }\href {https://doi.org/10.1103/PhysRevLett.99.146403} {\bibfield  {journal} {\bibinfo  {journal} {\prl}\ }\textbf {\bibinfo {volume} {99}},\ \bibinfo {pages} {146403}
  (\bibinfo {year} {2007})}\BibitemShut {NoStop}%
\bibitem [{\citenamefont {{Monney}}\ \emph {et~al.}(2011)\citenamefont {{Monney}}, \citenamefont {{Battaglia}}, \citenamefont {{Cercellier}}, \citenamefont {{Aebi}},\ and\ \citenamefont {{Beck}}}]{monney2011exciton}%
  \BibitemOpen
  \bibfield  {author} {\bibinfo {author} {\bibfnamefont {C.}~\bibnamefont {{Monney}}}, \bibinfo {author} {\bibfnamefont {C.}~\bibnamefont {{Battaglia}}}, \bibinfo {author} {\bibfnamefont {H.}~\bibnamefont {{Cercellier}}}, \bibinfo {author} {\bibfnamefont {P.}~\bibnamefont {{Aebi}}},\ and\ \bibinfo {author} {\bibfnamefont {H.}~\bibnamefont {{Beck}}},\ }\bibfield  {title} {\bibinfo {title} {{Exciton Condensation Driving the Periodic Lattice Distortion of 1T-TiSe$_{2}$}},\ }\href {https://doi.org/10.1103/PhysRevLett.106.106404} {\bibfield  {journal} {\bibinfo  {journal} {\prl}\ }\textbf {\bibinfo {volume} {106}},\ \bibinfo {pages} {106404} (\bibinfo {year} {2011})}\BibitemShut {NoStop}%
\bibitem [{\citenamefont {{Kogar}}\ \emph {et~al.}(2017)\citenamefont {{Kogar}}, \citenamefont {{Rak}}, \citenamefont {{Vig}}, \citenamefont {{Husain}}, \citenamefont {{Flicker}}, \citenamefont {{Joe}}, \citenamefont {{Venema}}, \citenamefont {{MacDougall}}, \citenamefont {{Chiang}}, \citenamefont {{Fradkin}}, \citenamefont {{van Wezel}},\ and\ \citenamefont {{Abbamonte}}}]{kogar2017signatures}%
  \BibitemOpen
  \bibfield  {author} {\bibinfo {author} {\bibfnamefont {A.}~\bibnamefont {{Kogar}}}, \bibinfo {author} {\bibfnamefont {M.~S.}\ \bibnamefont {{Rak}}}, \bibinfo {author} {\bibfnamefont {S.}~\bibnamefont {{Vig}}}, \bibinfo {author} {\bibfnamefont {A.~A.}\ \bibnamefont {{Husain}}}, \bibinfo {author} {\bibfnamefont {F.}~\bibnamefont {{Flicker}}}, \bibinfo {author} {\bibfnamefont {Y.~I.}\ \bibnamefont {{Joe}}}, \bibinfo {author} {\bibfnamefont {L.}~\bibnamefont {{Venema}}}, \bibinfo {author} {\bibfnamefont {G.~J.}\ \bibnamefont {{MacDougall}}}, \bibinfo {author} {\bibfnamefont {T.~C.}\ \bibnamefont {{Chiang}}}, \bibinfo {author} {\bibfnamefont {E.}~\bibnamefont {{Fradkin}}}, \bibinfo {author} {\bibfnamefont {J.}~\bibnamefont {{van Wezel}}},\ and\ \bibinfo {author} {\bibfnamefont {P.}~\bibnamefont {{Abbamonte}}},\ }\bibfield  {title} {\bibinfo {title} {{Signatures of exciton condensation in a transition metal dichalcogenide}},\ }\href {https://doi.org/10.1126/science.aam6432} {\bibfield  {journal} {\bibinfo
  {journal} {Science}\ }\textbf {\bibinfo {volume} {358}},\ \bibinfo {pages} {1314} (\bibinfo {year} {2017})}\BibitemShut {NoStop}%
\bibitem [{\citenamefont {{Wakisaka}}\ \emph {et~al.}(2009)\citenamefont {{Wakisaka}}, \citenamefont {{Sudayama}}, \citenamefont {{Takubo}}, \citenamefont {{Mizokawa}}, \citenamefont {{Arita}}, \citenamefont {{Namatame}}, \citenamefont {{Taniguchi}}, \citenamefont {{Katayama}}, \citenamefont {{Nohara}},\ and\ \citenamefont {{Takagi}}}]{wakisaka2009excitonic}%
  \BibitemOpen
  \bibfield  {author} {\bibinfo {author} {\bibfnamefont {Y.}~\bibnamefont {{Wakisaka}}}, \bibinfo {author} {\bibfnamefont {T.}~\bibnamefont {{Sudayama}}}, \bibinfo {author} {\bibfnamefont {K.}~\bibnamefont {{Takubo}}}, \bibinfo {author} {\bibfnamefont {T.}~\bibnamefont {{Mizokawa}}}, \bibinfo {author} {\bibfnamefont {M.}~\bibnamefont {{Arita}}}, \bibinfo {author} {\bibfnamefont {H.}~\bibnamefont {{Namatame}}}, \bibinfo {author} {\bibfnamefont {M.}~\bibnamefont {{Taniguchi}}}, \bibinfo {author} {\bibfnamefont {N.}~\bibnamefont {{Katayama}}}, \bibinfo {author} {\bibfnamefont {M.}~\bibnamefont {{Nohara}}},\ and\ \bibinfo {author} {\bibfnamefont {H.}~\bibnamefont {{Takagi}}},\ }\bibfield  {title} {\bibinfo {title} {{Excitonic Insulator State in Ta$_{2}$NiSe$_{5}$ Probed by Photoemission Spectroscopy}},\ }\href {https://doi.org/10.1103/PhysRevLett.103.026402} {\bibfield  {journal} {\bibinfo  {journal} {\prl}\ }\textbf {\bibinfo {volume} {103}},\ \bibinfo {pages} {026402} (\bibinfo {year} {2009})}\BibitemShut
  {NoStop}%
\bibitem [{\citenamefont {{Kaneko}}\ \emph {et~al.}(2013)\citenamefont {{Kaneko}}, \citenamefont {{Toriyama}}, \citenamefont {{Konishi}},\ and\ \citenamefont {{Ohta}}}]{kaneko2013orthorhombic}%
  \BibitemOpen
  \bibfield  {author} {\bibinfo {author} {\bibfnamefont {T.}~\bibnamefont {{Kaneko}}}, \bibinfo {author} {\bibfnamefont {T.}~\bibnamefont {{Toriyama}}}, \bibinfo {author} {\bibfnamefont {T.}~\bibnamefont {{Konishi}}},\ and\ \bibinfo {author} {\bibfnamefont {Y.}~\bibnamefont {{Ohta}}},\ }\bibfield  {title} {\bibinfo {title} {{Orthorhombic-to-monoclinic phase transition of Ta$_{2}$NiSe$_{5}$ induced by the Bose-Einstein condensation of excitons}},\ }\href {https://doi.org/10.1103/PhysRevB.87.035121} {\bibfield  {journal} {\bibinfo  {journal} {\prb}\ }\textbf {\bibinfo {volume} {87}},\ \bibinfo {pages} {035121} (\bibinfo {year} {2013})}\BibitemShut {NoStop}%
\bibitem [{\citenamefont {{Seki}}\ \emph {et~al.}(2014)\citenamefont {{Seki}}, \citenamefont {{Wakisaka}}, \citenamefont {{Kaneko}}, \citenamefont {{Toriyama}}, \citenamefont {{Konishi}}, \citenamefont {{Sudayama}}, \citenamefont {{Saini}}, \citenamefont {{Arita}}, \citenamefont {{Namatame}}, \citenamefont {{Taniguchi}}, \citenamefont {{Katayama}}, \citenamefont {{Nohara}}, \citenamefont {{Takagi}}, \citenamefont {{Mizokawa}},\ and\ \citenamefont {{Ohta}}}]{seki2014excitonic}%
  \BibitemOpen
  \bibfield  {author} {\bibinfo {author} {\bibfnamefont {K.}~\bibnamefont {{Seki}}}, \bibinfo {author} {\bibfnamefont {Y.}~\bibnamefont {{Wakisaka}}}, \bibinfo {author} {\bibfnamefont {T.}~\bibnamefont {{Kaneko}}}, \bibinfo {author} {\bibfnamefont {T.}~\bibnamefont {{Toriyama}}}, \bibinfo {author} {\bibfnamefont {T.}~\bibnamefont {{Konishi}}}, \bibinfo {author} {\bibfnamefont {T.}~\bibnamefont {{Sudayama}}}, \bibinfo {author} {\bibfnamefont {N.~L.}\ \bibnamefont {{Saini}}}, \bibinfo {author} {\bibfnamefont {M.}~\bibnamefont {{Arita}}}, \bibinfo {author} {\bibfnamefont {H.}~\bibnamefont {{Namatame}}}, \bibinfo {author} {\bibfnamefont {M.}~\bibnamefont {{Taniguchi}}}, \bibinfo {author} {\bibfnamefont {N.}~\bibnamefont {{Katayama}}}, \bibinfo {author} {\bibfnamefont {M.}~\bibnamefont {{Nohara}}}, \bibinfo {author} {\bibfnamefont {H.}~\bibnamefont {{Takagi}}}, \bibinfo {author} {\bibfnamefont {T.}~\bibnamefont {{Mizokawa}}},\ and\ \bibinfo {author} {\bibfnamefont {Y.}~\bibnamefont {{Ohta}}},\ }\bibfield  {title}
  {\bibinfo {title} {{Excitonic Bose-Einstein condensation in Ta$_{2}$NiSe$_{5}$ above room temperature}},\ }\href {https://doi.org/10.1103/PhysRevB.90.155116} {\bibfield  {journal} {\bibinfo  {journal} {\prb}\ }\textbf {\bibinfo {volume} {90}},\ \bibinfo {pages} {155116} (\bibinfo {year} {2014})}\BibitemShut {NoStop}%
\bibitem [{\citenamefont {{Bucher}}\ \emph {et~al.}(1991)\citenamefont {{Bucher}}, \citenamefont {{Steiner}},\ and\ \citenamefont {{Wachter}}}]{bucher1991excitonic}%
  \BibitemOpen
  \bibfield  {author} {\bibinfo {author} {\bibfnamefont {B.}~\bibnamefont {{Bucher}}}, \bibinfo {author} {\bibfnamefont {P.}~\bibnamefont {{Steiner}}},\ and\ \bibinfo {author} {\bibfnamefont {P.}~\bibnamefont {{Wachter}}},\ }\bibfield  {title} {\bibinfo {title} {{Excitonic insulator phase in TmSe$_{0.45}$Te$_{0.55}$}},\ }\href {https://doi.org/10.1103/PhysRevLett.67.2717} {\bibfield  {journal} {\bibinfo  {journal} {\prl}\ }\textbf {\bibinfo {volume} {67}},\ \bibinfo {pages} {2717} (\bibinfo {year} {1991})}\BibitemShut {NoStop}%
\bibitem [{\citenamefont {{Zhang}}\ \emph {et~al.}(2024)\citenamefont {{Zhang}}, \citenamefont {{Dong}}, \citenamefont {{Yan}}, \citenamefont {{Jiang}}, \citenamefont {{Yang}}, \citenamefont {{Li}}, \citenamefont {{Guo}}, \citenamefont {{Huang}}, \citenamefont {{Haobo}}, \citenamefont {{Li}}, \citenamefont {{Kurokawa}}, \citenamefont {{Wang}}, \citenamefont {{Nie}}, \citenamefont {{Hashimoto}}, \citenamefont {{Lu}}, \citenamefont {{Jiao}}, \citenamefont {{Shen}}, \citenamefont {{Qian}}, \citenamefont {{Wang}}, \citenamefont {{Shi}},\ and\ \citenamefont {{Kondo}}}]{Zhang2024spontaneous}%
  \BibitemOpen
  \bibfield  {author} {\bibinfo {author} {\bibfnamefont {P.}~\bibnamefont {{Zhang}}}, \bibinfo {author} {\bibfnamefont {Y.}~\bibnamefont {{Dong}}}, \bibinfo {author} {\bibfnamefont {D.}~\bibnamefont {{Yan}}}, \bibinfo {author} {\bibfnamefont {B.}~\bibnamefont {{Jiang}}}, \bibinfo {author} {\bibfnamefont {T.}~\bibnamefont {{Yang}}}, \bibinfo {author} {\bibfnamefont {J.}~\bibnamefont {{Li}}}, \bibinfo {author} {\bibfnamefont {Z.}~\bibnamefont {{Guo}}}, \bibinfo {author} {\bibfnamefont {Y.}~\bibnamefont {{Huang}}}, \bibinfo {author} {\bibfnamefont {L.}~\bibnamefont {{Haobo}}, \bibfnamefont {Qing}}, \bibinfo {author} {\bibfnamefont {Y.}~\bibnamefont {{Li}}}, \bibinfo {author} {\bibfnamefont {K.}~\bibnamefont {{Kurokawa}}}, \bibinfo {author} {\bibfnamefont {R.}~\bibnamefont {{Wang}}}, \bibinfo {author} {\bibfnamefont {Y.}~\bibnamefont {{Nie}}}, \bibinfo {author} {\bibfnamefont {M.}~\bibnamefont {{Hashimoto}}}, \bibinfo {author} {\bibfnamefont {D.}~\bibnamefont {{Lu}}}, \bibinfo {author} {\bibfnamefont {W.-H.}\
  \bibnamefont {{Jiao}}}, \bibinfo {author} {\bibfnamefont {J.}~\bibnamefont {{Shen}}}, \bibinfo {author} {\bibfnamefont {T.}~\bibnamefont {{Qian}}}, \bibinfo {author} {\bibfnamefont {Z.}~\bibnamefont {{Wang}}}, \bibinfo {author} {\bibfnamefont {Y.}~\bibnamefont {{Shi}}},\ and\ \bibinfo {author} {\bibfnamefont {T.}~\bibnamefont {{Kondo}}},\ }\bibfield  {title} {\bibinfo {title} {{Spontaneous Gap Opening and Potential Excitonic States in an Ideal Dirac Semimetal Ta$_{2}$Pd$_{3}$Te$_{5}$}},\ }\href {https://doi.org/10.1103/PhysRevX.14.011047} {\bibfield  {journal} {\bibinfo  {journal} {Phys, Rev. X}\ }\textbf {\bibinfo {volume} {14}},\ \bibinfo {pages} {011047} (\bibinfo {year} {2024})}\BibitemShut {NoStop}%
\bibitem [{\citenamefont {{Huang}}\ \emph {et~al.}(2024)\citenamefont {{Huang}}, \citenamefont {{Jiang}}, \citenamefont {{Yao}}, \citenamefont {{Yan}}, \citenamefont {{Lei}}, \citenamefont {{Gao}}, \citenamefont {{Guo}}, \citenamefont {{Jin}}, \citenamefont {{Li}}, \citenamefont {{Yuan}}, \citenamefont {{Chai}}, \citenamefont {{Sheng}}, \citenamefont {{Pan}}, \citenamefont {{Chen}}, \citenamefont {{Liu}}, \citenamefont {{Gao}}, \citenamefont {{Qu}}, \citenamefont {{Liu}}, \citenamefont {{Jiang}}, \citenamefont {{Liu}}, \citenamefont {{Ma}}, \citenamefont {{Zhou}}, \citenamefont {{Huang}}, \citenamefont {{Yun}}, \citenamefont {{Zhang}}, \citenamefont {{Li}}, \citenamefont {{Jin}}, \citenamefont {{Ding}}, \citenamefont {{Shen}}, \citenamefont {{Su}}, \citenamefont {{Shi}}, \citenamefont {{Wang}},\ and\ \citenamefont {{Qian}}}]{huang2024evidence}%
  \BibitemOpen
  \bibfield  {author} {\bibinfo {author} {\bibfnamefont {J.}~\bibnamefont {{Huang}}}, \bibinfo {author} {\bibfnamefont {B.}~\bibnamefont {{Jiang}}}, \bibinfo {author} {\bibfnamefont {J.}~\bibnamefont {{Yao}}}, \bibinfo {author} {\bibfnamefont {D.}~\bibnamefont {{Yan}}}, \bibinfo {author} {\bibfnamefont {X.}~\bibnamefont {{Lei}}}, \bibinfo {author} {\bibfnamefont {J.}~\bibnamefont {{Gao}}}, \bibinfo {author} {\bibfnamefont {Z.}~\bibnamefont {{Guo}}}, \bibinfo {author} {\bibfnamefont {F.}~\bibnamefont {{Jin}}}, \bibinfo {author} {\bibfnamefont {Y.}~\bibnamefont {{Li}}}, \bibinfo {author} {\bibfnamefont {Z.}~\bibnamefont {{Yuan}}}, \bibinfo {author} {\bibfnamefont {C.}~\bibnamefont {{Chai}}}, \bibinfo {author} {\bibfnamefont {H.}~\bibnamefont {{Sheng}}}, \bibinfo {author} {\bibfnamefont {M.}~\bibnamefont {{Pan}}}, \bibinfo {author} {\bibfnamefont {F.}~\bibnamefont {{Chen}}}, \bibinfo {author} {\bibfnamefont {J.}~\bibnamefont {{Liu}}}, \bibinfo {author} {\bibfnamefont {S.}~\bibnamefont {{Gao}}}, \bibinfo {author}
  {\bibfnamefont {G.}~\bibnamefont {{Qu}}}, \bibinfo {author} {\bibfnamefont {B.}~\bibnamefont {{Liu}}}, \bibinfo {author} {\bibfnamefont {Z.}~\bibnamefont {{Jiang}}}, \bibinfo {author} {\bibfnamefont {Z.}~\bibnamefont {{Liu}}}, \bibinfo {author} {\bibfnamefont {X.}~\bibnamefont {{Ma}}}, \bibinfo {author} {\bibfnamefont {S.}~\bibnamefont {{Zhou}}}, \bibinfo {author} {\bibfnamefont {Y.}~\bibnamefont {{Huang}}}, \bibinfo {author} {\bibfnamefont {C.}~\bibnamefont {{Yun}}}, \bibinfo {author} {\bibfnamefont {Q.}~\bibnamefont {{Zhang}}}, \bibinfo {author} {\bibfnamefont {S.}~\bibnamefont {{Li}}}, \bibinfo {author} {\bibfnamefont {S.}~\bibnamefont {{Jin}}}, \bibinfo {author} {\bibfnamefont {H.}~\bibnamefont {{Ding}}}, \bibinfo {author} {\bibfnamefont {J.}~\bibnamefont {{Shen}}}, \bibinfo {author} {\bibfnamefont {D.}~\bibnamefont {{Su}}}, \bibinfo {author} {\bibfnamefont {Y.}~\bibnamefont {{Shi}}}, \bibinfo {author} {\bibfnamefont {Z.}~\bibnamefont {{Wang}}},\ and\ \bibinfo {author} {\bibfnamefont {T.}~\bibnamefont
  {{Qian}}},\ }\bibfield  {title} {\bibinfo {title} {{Evidence for an Excitonic Insulator State in Ta$_{2}$Pd$_{3}$Te$_{5}$}},\ }\href {https://doi.org/10.1103/PhysRevX.14.011046} {\bibfield  {journal} {\bibinfo  {journal} {Phys. Rev. X}\ }\textbf {\bibinfo {volume} {14}},\ \bibinfo {pages} {011046} (\bibinfo {year} {2024})}\BibitemShut {NoStop}%
  

\bibitem [{\citenamefont {{Shafayat Hossain}}\ \emph {et~al.}(2025)\citenamefont {{Shafayat Hossain}}, \citenamefont {{Cochran}}, \citenamefont {{Jiang}}, \citenamefont {{Zhang}}, \citenamefont {{Wu}}, \citenamefont {{Liu}}, \citenamefont {{Zheng}}, \citenamefont {{Kim}}, \citenamefont {{Cheng}}, \citenamefont {{Zhang}}, \citenamefont {{Litskevich}}, \citenamefont {{Zhang}}, \citenamefont {{Cheng}}, \citenamefont {{Liu}}, \citenamefont {{Yin}}, \citenamefont {{Yang}}, \citenamefont {{Denlinger}}, \citenamefont {{Tallarida}}, \citenamefont {{Dai}}, \citenamefont {{Vescovo}}, \citenamefont {{Rajapitamahuni}}, \citenamefont {{Miao}}, \citenamefont {{Yao}}, \citenamefont {{Keselman}}, \citenamefont {{Peng}}, \citenamefont {{Yao}}, \citenamefont {{Wang}}, \citenamefont {{Balicas}}, \citenamefont {{Neupert}},\ and\ \citenamefont {{Zahid Hasan}}}]{hossain2025topological}%
  \BibitemOpen
  \bibfield  {author} {\bibinfo {author} {\bibfnamefont {M.}~\bibnamefont {{Shafayat Hossain}}}, \bibinfo {author} {\bibfnamefont {T.~A.}\ \bibnamefont {{Cochran}}}, \bibinfo {author} {\bibfnamefont {Y.-X.}\ \bibnamefont {{Jiang}}}, \bibinfo {author} {\bibfnamefont {S.}~\bibnamefont {{Zhang}}}, \bibinfo {author} {\bibfnamefont {H.}~\bibnamefont {{Wu}}}, \bibinfo {author} {\bibfnamefont {X.}~\bibnamefont {{Liu}}}, \bibinfo {author} {\bibfnamefont {X.}~\bibnamefont {{Zheng}}}, \bibinfo {author} {\bibfnamefont {B.}~\bibnamefont {{Kim}}}, \bibinfo {author} {\bibfnamefont {G.}~\bibnamefont {{Cheng}}}, \bibinfo {author} {\bibfnamefont {Q.}~\bibnamefont {{Zhang}}}, \bibinfo {author} {\bibfnamefont {M.}~\bibnamefont {{Litskevich}}}, \bibinfo {author} {\bibfnamefont {J.}~\bibnamefont {{Zhang}}}, \bibinfo {author} {\bibfnamefont {Z.-J.}\ \bibnamefont {{Cheng}}}, \bibinfo {author} {\bibfnamefont {J.}~\bibnamefont {{Liu}}}, \bibinfo {author} {\bibfnamefont {J.-X.}\ \bibnamefont {{Yin}}}, \bibinfo {author} {\bibfnamefont
  {X.~P.}\ \bibnamefont {{Yang}}}, \bibinfo {author} {\bibfnamefont {J.}~\bibnamefont {{Denlinger}}}, \bibinfo {author} {\bibfnamefont {M.}~\bibnamefont {{Tallarida}}}, \bibinfo {author} {\bibfnamefont {J.}~\bibnamefont {{Dai}}}, \bibinfo {author} {\bibfnamefont {E.}~\bibnamefont {{Vescovo}}}, \bibinfo {author} {\bibfnamefont {A.}~\bibnamefont {{Rajapitamahuni}}}, \bibinfo {author} {\bibfnamefont {H.}~\bibnamefont {{Miao}}}, \bibinfo {author} {\bibfnamefont {N.}~\bibnamefont {{Yao}}}, \bibinfo {author} {\bibfnamefont {A.}~\bibnamefont {{Keselman}}}, \bibinfo {author} {\bibfnamefont {Y.}~\bibnamefont {{Peng}}}, \bibinfo {author} {\bibfnamefont {Y.}~\bibnamefont {{Yao}}}, \bibinfo {author} {\bibfnamefont {Z.}~\bibnamefont {{Wang}}}, \bibinfo {author} {\bibfnamefont {L.}~\bibnamefont {{Balicas}}}, \bibinfo {author} {\bibfnamefont {T.}~\bibnamefont {{Neupert}}},\ and\ \bibinfo {author} {\bibfnamefont {M.}~\bibnamefont {{Zahid Hasan}}},\ }\bibfield  {title} {\bibinfo {title} {Topological excitonic insulator with
  tunable momentum order},\ }\href {https://doi.org/https://doi.org/10.1038/s41567-025-02917-6} {\bibfield  {journal} {\bibinfo  {journal} {Nat. Phys.}\ }\textbf {\bibinfo {volume} {21}},\ \bibinfo {pages} {1} (\bibinfo {year} {2025})}\BibitemShut {NoStop}%
\bibitem [{\citenamefont {{Liimatta}}\ and\ \citenamefont {{Ibers}}(1987)}]{sunshine1985structure}%
  \BibitemOpen
  \bibfield  {author} {\bibinfo {author} {\bibfnamefont {E.~W.}\ \bibnamefont {{Liimatta}}}\ and\ \bibinfo {author} {\bibfnamefont {J.~A.}\ \bibnamefont {{Ibers}}},\ }\bibfield  {title} {\bibinfo {title} {{Synthesis, structure, and physical properties of the new layered ternary chalcogenide NbNiTe$_{5}$}},\ }\href {https://doi.org/10.1016/0022-4596(87)90246-5} {\bibfield  {journal} {\bibinfo  {journal} {J. Solid State Chem. France}\ }\textbf {\bibinfo {volume} {71}},\ \bibinfo {pages} {384} (\bibinfo {year} {1987})}\BibitemShut {NoStop}%
\bibitem [{\citenamefont {{Klein}}\ \emph {et~al.}(2013)\citenamefont {{Klein}}, \citenamefont {{Moutaabbid}}, \citenamefont {{Boubiche}}, \citenamefont {{Soyer}}, \citenamefont {{Rousse}},\ and\ \citenamefont {{Gauzzi}}}]{di1986physical}%
  \BibitemOpen
  \bibfield  {author} {\bibinfo {author} {\bibfnamefont {Y.}~\bibnamefont {{Klein}}}, \bibinfo {author} {\bibfnamefont {H.}~\bibnamefont {{Moutaabbid}}}, \bibinfo {author} {\bibfnamefont {N.}~\bibnamefont {{Boubiche}}}, \bibinfo {author} {\bibfnamefont {A.}~\bibnamefont {{Soyer}}}, \bibinfo {author} {\bibfnamefont {G.}~\bibnamefont {{Rousse}}},\ and\ \bibinfo {author} {\bibfnamefont {A.}~\bibnamefont {{Gauzzi}}},\ }\bibfield  {title} {\bibinfo {title} {{Weak localization and Mott state in two-dimensional Sr$_{3}$V$_{5}$S$_{11}$}},\ }\href {https://doi.org/10.1088/0953-8984/25/43/435604} {\bibfield  {journal} {\bibinfo  {journal} {J. Phys. Condens.}\ }\textbf {\bibinfo {volume} {25}},\ \bibinfo {pages} {435604} (\bibinfo {year} {2013})}\BibitemShut {NoStop}%
\bibitem [{\citenamefont {{Kim}}\ \emph {et~al.}(2016)\citenamefont {{Kim}}, \citenamefont {{Kim}}, \citenamefont {{Kang}}, \citenamefont {{An}}, \citenamefont {{Kim}}, \citenamefont {{Eom}}, \citenamefont {{Lee}}, \citenamefont {{Park}}, \citenamefont {{Kim}}, \citenamefont {{Choi}}, \citenamefont {{Min}},\ and\ \citenamefont {{Kim}}}]{kim2016layer}%
  \BibitemOpen
  \bibfield  {author} {\bibinfo {author} {\bibfnamefont {S.~Y.}\ \bibnamefont {{Kim}}}, \bibinfo {author} {\bibfnamefont {Y.}~\bibnamefont {{Kim}}}, \bibinfo {author} {\bibfnamefont {C.}~\bibnamefont {{Kang}}}, \bibinfo {author} {\bibfnamefont {E.}~\bibnamefont {{An}}}, \bibinfo {author} {\bibfnamefont {H.~K.}\ \bibnamefont {{Kim}}}, \bibinfo {author} {\bibfnamefont {M.~J.}\ \bibnamefont {{Eom}}}, \bibinfo {author} {\bibfnamefont {M.}~\bibnamefont {{Lee}}}, \bibinfo {author} {\bibfnamefont {C.}~\bibnamefont {{Park}}}, \bibinfo {author} {\bibfnamefont {T.}~\bibnamefont {{Kim}}}, \bibinfo {author} {\bibfnamefont {H.}~\bibnamefont {{Choi}}}, \bibinfo {author} {\bibfnamefont {B.~I.}\ \bibnamefont {{Min}}},\ and\ \bibinfo {author} {\bibfnamefont {J.~S.}\ \bibnamefont {{Kim}}},\ }\bibfield  {title} {\bibinfo {title} {Layer-confined excitonic insulating phase in ultrathin {Ta$_2$NiSe$_5$} crystals},\ }\href {https://doi.org/10.1021/acsnano.6b04796} {\bibfield  {journal} {\bibinfo  {journal} {ACS Nano}\ }\textbf
  {\bibinfo {volume} {10}},\ \bibinfo {pages} {8888} (\bibinfo {year} {2016})}\BibitemShut {NoStop}%
\bibitem [{\citenamefont {{Chen}}\ \emph {et~al.}(2023{\natexlab{a}})\citenamefont {{Chen}}, \citenamefont {{Chen}}, \citenamefont {{Tang}}, \citenamefont {{Li}}, \citenamefont {{Wang}}, \citenamefont {{Ding}}, \citenamefont {{Kang}}, \citenamefont {{Jozwiak}}, \citenamefont {{Bostwick}}, \citenamefont {{Rotenberg}}, \citenamefont {{Hashimoto}}, \citenamefont {{Lu}}, \citenamefont {{Ruff}}, \citenamefont {{Louie}}, \citenamefont {{Birgeneau}}, \citenamefont {{Chen}}, \citenamefont {{Wang}},\ and\ \citenamefont {{He}}}]{chen2023role}%
  \BibitemOpen
  \bibfield  {author} {\bibinfo {author} {\bibfnamefont {C.}~\bibnamefont {{Chen}}}, \bibinfo {author} {\bibfnamefont {X.}~\bibnamefont {{Chen}}}, \bibinfo {author} {\bibfnamefont {W.}~\bibnamefont {{Tang}}}, \bibinfo {author} {\bibfnamefont {Z.}~\bibnamefont {{Li}}}, \bibinfo {author} {\bibfnamefont {S.}~\bibnamefont {{Wang}}}, \bibinfo {author} {\bibfnamefont {S.}~\bibnamefont {{Ding}}}, \bibinfo {author} {\bibfnamefont {Z.}~\bibnamefont {{Kang}}}, \bibinfo {author} {\bibfnamefont {C.}~\bibnamefont {{Jozwiak}}}, \bibinfo {author} {\bibfnamefont {A.}~\bibnamefont {{Bostwick}}}, \bibinfo {author} {\bibfnamefont {E.}~\bibnamefont {{Rotenberg}}}, \bibinfo {author} {\bibfnamefont {M.}~\bibnamefont {{Hashimoto}}}, \bibinfo {author} {\bibfnamefont {D.}~\bibnamefont {{Lu}}}, \bibinfo {author} {\bibfnamefont {J.~P.~C.}\ \bibnamefont {{Ruff}}}, \bibinfo {author} {\bibfnamefont {S.~G.}\ \bibnamefont {{Louie}}}, \bibinfo {author} {\bibfnamefont {R.~J.}\ \bibnamefont {{Birgeneau}}}, \bibinfo {author} {\bibfnamefont
  {Y.}~\bibnamefont {{Chen}}}, \bibinfo {author} {\bibfnamefont {Y.}~\bibnamefont {{Wang}}},\ and\ \bibinfo {author} {\bibfnamefont {Y.}~\bibnamefont {{He}}},\ }\bibfield  {title} {\bibinfo {title} {{Role of electron-phonon coupling in excitonic insulator candidate Ta$_{2}$NiSe$_{5}$}},\ }\href {https://doi.org/10.1103/PhysRevResearch.5.043089} {\bibfield  {journal} {\bibinfo  {journal} {Phys. Rev. Res.}\ }\textbf {\bibinfo {volume} {5}},\ \bibinfo {pages} {043089} (\bibinfo {year} {2023}{\natexlab{a}})}\BibitemShut {NoStop}%
\bibitem [{\citenamefont {{Nakano}}\ \emph {et~al.}(2018)\citenamefont {{Nakano}}, \citenamefont {{Hasegawa}}, \citenamefont {{Tamura}}, \citenamefont {{Katayama}}, \citenamefont {{Tsutsui}},\ and\ \citenamefont {{Sawa}}}]{nakano2018antiferroelectric}%
  \BibitemOpen
  \bibfield  {author} {\bibinfo {author} {\bibfnamefont {A.}~\bibnamefont {{Nakano}}}, \bibinfo {author} {\bibfnamefont {T.}~\bibnamefont {{Hasegawa}}}, \bibinfo {author} {\bibfnamefont {S.}~\bibnamefont {{Tamura}}}, \bibinfo {author} {\bibfnamefont {N.}~\bibnamefont {{Katayama}}}, \bibinfo {author} {\bibfnamefont {S.}~\bibnamefont {{Tsutsui}}},\ and\ \bibinfo {author} {\bibfnamefont {H.}~\bibnamefont {{Sawa}}},\ }\bibfield  {title} {\bibinfo {title} {{Antiferroelectric distortion with anomalous phonon softening in the excitonic insulator Ta$_{2}$NiSe$_{5}$}},\ }\href {https://doi.org/10.1103/PhysRevB.98.045139} {\bibfield  {journal} {\bibinfo  {journal} {\prb}\ }\textbf {\bibinfo {volume} {98}},\ \bibinfo {pages} {045139} (\bibinfo {year} {2018})}\BibitemShut {NoStop}%
\bibitem [{\citenamefont {{Mazza}}\ \emph {et~al.}(2020)\citenamefont {{Mazza}}, \citenamefont {{R{\"o}sner}}, \citenamefont {{Windg{\"a}tter}}, \citenamefont {{Latini}}, \citenamefont {{H{\"u}bener}}, \citenamefont {{Millis}}, \citenamefont {{Rubio}},\ and\ \citenamefont {{Georges}}}]{mazza2020nature}%
  \BibitemOpen
  \bibfield  {author} {\bibinfo {author} {\bibfnamefont {G.}~\bibnamefont {{Mazza}}}, \bibinfo {author} {\bibfnamefont {M.}~\bibnamefont {{R{\"o}sner}}}, \bibinfo {author} {\bibfnamefont {L.}~\bibnamefont {{Windg{\"a}tter}}}, \bibinfo {author} {\bibfnamefont {S.}~\bibnamefont {{Latini}}}, \bibinfo {author} {\bibfnamefont {H.}~\bibnamefont {{H{\"u}bener}}}, \bibinfo {author} {\bibfnamefont {A.~J.}\ \bibnamefont {{Millis}}}, \bibinfo {author} {\bibfnamefont {A.}~\bibnamefont {{Rubio}}},\ and\ \bibinfo {author} {\bibfnamefont {A.}~\bibnamefont {{Georges}}},\ }\bibfield  {title} {\bibinfo {title} {{Nature of Symmetry Breaking at the Excitonic Insulator Transition: Ta$_{2}$NiSe$_{5}$}},\ }\href {https://doi.org/10.1103/PhysRevLett.124.197601} {\bibfield  {journal} {\bibinfo  {journal} {\prl}\ }\textbf {\bibinfo {volume} {124}},\ \bibinfo {pages} {197601} (\bibinfo {year} {2020})}\BibitemShut {NoStop}%
\bibitem [{Sup()}]{SupplementalMaterial}%
  \BibitemOpen
  \href {http://link.aps.org/ supplemental/} {\bibinfo {title} {http://link.aps.org/ supplemental/}}\BibitemShut {NoStop}%
\bibitem [{\citenamefont {Bie}\ \emph {et~al.}(2021)\citenamefont {Bie}, \citenamefont {Zong}, \citenamefont {Wang}, \citenamefont {Jarillo-Herrero},\ and\ \citenamefont {Gedik}}]{Bie2021}%
  \BibitemOpen
  \bibfield  {author} {\bibinfo {author} {\bibfnamefont {Y.-Q.}\ \bibnamefont {Bie}}, \bibinfo {author} {\bibfnamefont {A.}~\bibnamefont {Zong}}, \bibinfo {author} {\bibfnamefont {X.}~\bibnamefont {Wang}}, \bibinfo {author} {\bibfnamefont {P.}~\bibnamefont {Jarillo-Herrero}},\ and\ \bibinfo {author} {\bibfnamefont {N.}~\bibnamefont {Gedik}},\ }\bibfield  {title} {\bibinfo {title} {A versatile sample fabrication method for ultrafast electron diffraction},\ }\href {https://doi.org/10.1016/j.ultramic.2021.113389} {\bibfield  {journal} {\bibinfo  {journal} {Ultramicroscopy}\ }\textbf {\bibinfo {volume} {230}},\ \bibinfo {pages} {113389} (\bibinfo {year} {2021})}\BibitemShut {NoStop}%
\bibitem [{\citenamefont {Cheng}\ \emph {et~al.}(2024)\citenamefont {Cheng}, \citenamefont {Zong}, \citenamefont {Wu}, \citenamefont {Meng}, \citenamefont {Xia}, \citenamefont {Qi}, \citenamefont {Zhu}, \citenamefont {Zou}, \citenamefont {Jiang}, \citenamefont {Guo}, \citenamefont {van Wezel}, \citenamefont {Kogar}, \citenamefont {Zuerch}, \citenamefont {Zhang}, \citenamefont {Zhu},\ and\ \citenamefont {Xiang}}]{Cheng2024ultrafast}%
  \BibitemOpen
  \bibfield  {author} {\bibinfo {author} {\bibfnamefont {Y.}~\bibnamefont {Cheng}}, \bibinfo {author} {\bibfnamefont {A.}~\bibnamefont {Zong}}, \bibinfo {author} {\bibfnamefont {L.}~\bibnamefont {Wu}}, \bibinfo {author} {\bibfnamefont {Q.}~\bibnamefont {Meng}}, \bibinfo {author} {\bibfnamefont {W.}~\bibnamefont {Xia}}, \bibinfo {author} {\bibfnamefont {F.}~\bibnamefont {Qi}}, \bibinfo {author} {\bibfnamefont {P.}~\bibnamefont {Zhu}}, \bibinfo {author} {\bibfnamefont {X.}~\bibnamefont {Zou}}, \bibinfo {author} {\bibfnamefont {T.}~\bibnamefont {Jiang}}, \bibinfo {author} {\bibfnamefont {Y.}~\bibnamefont {Guo}}, \bibinfo {author} {\bibfnamefont {J.}~\bibnamefont {van Wezel}}, \bibinfo {author} {\bibfnamefont {A.}~\bibnamefont {Kogar}}, \bibinfo {author} {\bibfnamefont {M.~W.}\ \bibnamefont {Zuerch}}, \bibinfo {author} {\bibfnamefont {J.}~\bibnamefont {Zhang}}, \bibinfo {author} {\bibfnamefont {Y.}~\bibnamefont {Zhu}},\ and\ \bibinfo {author} {\bibfnamefont {D.}~\bibnamefont {Xiang}},\ }\bibfield  {title}
  {\bibinfo {title} {Ultrafast formation of topological defects in a two-dimensional charge density wave},\ }\href {https://doi.org/10.1038/s41567-023-02279-x} {\bibfield  {journal} {\bibinfo  {journal} {Nat. Phys.}\ }\textbf {\bibinfo {volume} {20}},\ \bibinfo {pages} {54} (\bibinfo {year} {2024})}\BibitemShut {NoStop}%
\bibitem [{\citenamefont {Zong}(2021)}]{zong2021emergent}%
  \BibitemOpen
  \bibfield  {author} {\bibinfo {author} {\bibfnamefont {A.}~\bibnamefont {Zong}},\ }\href {https://doi.org/10.1007/978-3-030-81751-0} {\emph {\bibinfo {title} {{Emergent States in Photoinduced Charge-Density-Wave Transitions}}}}\ (\bibinfo  {publisher} {Springer Nature},\ \bibinfo {year} {2021})\BibitemShut {NoStop}%
\bibitem [{\citenamefont {{Seo}}\ \emph {et~al.}(2018)\citenamefont {{Seo}}, \citenamefont {{Eom}}, \citenamefont {{Kim}}, \citenamefont {{Kang}}, \citenamefont {{Min}},\ and\ \citenamefont {{Hwang}}}]{seo2018temperarue}%
  \BibitemOpen
  \bibfield  {author} {\bibinfo {author} {\bibfnamefont {Y.-S.}\ \bibnamefont {{Seo}}}, \bibinfo {author} {\bibfnamefont {M.~J.}\ \bibnamefont {{Eom}}}, \bibinfo {author} {\bibfnamefont {J.~S.}\ \bibnamefont {{Kim}}}, \bibinfo {author} {\bibfnamefont {C.-J.}\ \bibnamefont {{Kang}}}, \bibinfo {author} {\bibfnamefont {B.~I.}\ \bibnamefont {{Min}}},\ and\ \bibinfo {author} {\bibfnamefont {J.}~\bibnamefont {{Hwang}}},\ }\bibfield  {title} {\bibinfo {title} {{{Temperature-dependent excitonic superfluid plasma frequency evolution in an excitonic insulator, Ta$_{2}$NiSe$_{5}$}}},\ }\href {https://doi.org/10.1038/s41598-018-30430-9} {\bibfield  {journal} {\bibinfo  {journal} {Sci. Rep.}\ }\textbf {\bibinfo {volume} {8}},\ \bibinfo {pages} {11961} (\bibinfo {year} {2018})}\BibitemShut {NoStop}%
\bibitem [{\citenamefont {{Lu}}\ \emph {et~al.}(2017)\citenamefont {{Lu}}, \citenamefont {{Kono}}, \citenamefont {{Larkin}}, \citenamefont {{Rost}}, \citenamefont {{Takayama}}, \citenamefont {{Boris}}, \citenamefont {{Keimer}},\ and\ \citenamefont {{Takagi}}}]{lu2017zerogap}%
  \BibitemOpen
  \bibfield  {author} {\bibinfo {author} {\bibfnamefont {Y.~F.}\ \bibnamefont {{Lu}}}, \bibinfo {author} {\bibfnamefont {H.}~\bibnamefont {{Kono}}}, \bibinfo {author} {\bibfnamefont {T.~I.}\ \bibnamefont {{Larkin}}}, \bibinfo {author} {\bibfnamefont {A.~W.}\ \bibnamefont {{Rost}}}, \bibinfo {author} {\bibfnamefont {T.}~\bibnamefont {{Takayama}}}, \bibinfo {author} {\bibfnamefont {A.~V.}\ \bibnamefont {{Boris}}}, \bibinfo {author} {\bibfnamefont {B.}~\bibnamefont {{Keimer}}},\ and\ \bibinfo {author} {\bibfnamefont {H.}~\bibnamefont {{Takagi}}},\ }\bibfield  {title} {\bibinfo {title} {{{Zero-gap semiconductor to excitonic insulator transition in Ta$_{2}$NiSe$_{5}$}}},\ }\href {https://doi.org/10.1038/ncomms14408} {\bibfield  {journal} {\bibinfo  {journal} {Nat. Commun.}\ }\textbf {\bibinfo {volume} {8}},\ \bibinfo {pages} {14408} (\bibinfo {year} {2017})}\BibitemShut {NoStop}%
\bibitem [{\citenamefont {Kirkland}(1998)}]{kirkland1998advanced}%
  \BibitemOpen
  \bibfield  {author} {\bibinfo {author} {\bibfnamefont {E.~J.}\ \bibnamefont {Kirkland}},\ }\href {https://doi.org/10.1007/978-1-4419-6533-2} {\emph {\bibinfo {title} {Advanced computing in electron microscopy}}},\ Vol.~\bibinfo {volume} {12}\ (\bibinfo  {publisher} {Springer},\ \bibinfo {year} {1998})\BibitemShut {NoStop}%
\bibitem [{\citenamefont {{Furness}}\ \emph {et~al.}(2020)\citenamefont {{Furness}}, \citenamefont {{Kaplan}}, \citenamefont {{Ning}}, \citenamefont {{Perdew}},\ and\ \citenamefont {{Sun}}}]{furness2020accurate}%
  \BibitemOpen
  \bibfield  {author} {\bibinfo {author} {\bibfnamefont {J.~W.}\ \bibnamefont {{Furness}}}, \bibinfo {author} {\bibfnamefont {A.~D.}\ \bibnamefont {{Kaplan}}}, \bibinfo {author} {\bibfnamefont {J.}~\bibnamefont {{Ning}}}, \bibinfo {author} {\bibfnamefont {J.~P.}\ \bibnamefont {{Perdew}}},\ and\ \bibinfo {author} {\bibfnamefont {J.}~\bibnamefont {{Sun}}},\ }\bibfield  {title} {\bibinfo {title} {{Accurate and numerically efficient r$^2$SCAN meta-generalized gradient approximation}},\ }\href {https://doi.org/10.1021/acs.jpclett.0c02405} {\bibfield  {journal} {\bibinfo  {journal} {J. Phys. Chem. Lett.}\ }\textbf {\bibinfo {volume} {11}},\ \bibinfo {pages} {8208} (\bibinfo {year} {2020})}\BibitemShut {NoStop}%
\bibitem [{\citenamefont {{Giannozzi}}\ \emph {et~al.}(2017)\citenamefont {{Giannozzi}}, \citenamefont {{Andreussi}}, \citenamefont {{Brumme}}, \citenamefont {{Bunau}}, \citenamefont {{Buongiorno Nardelli}}, \citenamefont {{Calandra}}, \citenamefont {{Car}}, \citenamefont {{Cavazzoni}}, \citenamefont {{Ceresoli}}, \citenamefont {{Cococcioni}}, \citenamefont {{Colonna}}, \citenamefont {{Carnimeo}}, \citenamefont {{Dal Corso}}, \citenamefont {{de Gironcoli}}, \citenamefont {{Delugas}}, \citenamefont {{DiStasio}}, \citenamefont {{Ferretti}}, \citenamefont {{Floris}}, \citenamefont {{Fratesi}}, \citenamefont {{Fugallo}}, \citenamefont {{Gebauer}}, \citenamefont {{Gerstmann}}, \citenamefont {{Giustino}}, \citenamefont {{Gorni}}, \citenamefont {{Jia}}, \citenamefont {{Kawamura}}, \citenamefont {{Ko}}, \citenamefont {{Kokalj}}, \citenamefont {{K{\"u}{\c{c}}{\"u}kbenli}}, \citenamefont {{Lazzeri}}, \citenamefont {{Marsili}}, \citenamefont {{Marzari}}, \citenamefont {{Mauri}}, \citenamefont {{Nguyen}}, \citenamefont
  {{Nguyen}}, \citenamefont {{Otero-de-la-Roza}}, \citenamefont {{Paulatto}}, \citenamefont {{Ponc{\'e}}}, \citenamefont {{Rocca}}, \citenamefont {{Sabatini}}, \citenamefont {{Santra}}, \citenamefont {{Schlipf}}, \citenamefont {{Seitsonen}}, \citenamefont {{Smogunov}}, \citenamefont {{Timrov}}, \citenamefont {{Thonhauser}}, \citenamefont {{Umari}}, \citenamefont {{Vast}}, \citenamefont {{Wu}},\ and\ \citenamefont {{Baroni}}}]{giannozzi2017advanced}%
  \BibitemOpen
  \bibfield  {author} {\bibinfo {author} {\bibfnamefont {P.}~\bibnamefont {{Giannozzi}}}, \bibinfo {author} {\bibfnamefont {O.}~\bibnamefont {{Andreussi}}}, \bibinfo {author} {\bibfnamefont {T.}~\bibnamefont {{Brumme}}}, \bibinfo {author} {\bibfnamefont {O.}~\bibnamefont {{Bunau}}}, \bibinfo {author} {\bibfnamefont {M.}~\bibnamefont {{Buongiorno Nardelli}}}, \bibinfo {author} {\bibfnamefont {M.}~\bibnamefont {{Calandra}}}, \bibinfo {author} {\bibfnamefont {R.}~\bibnamefont {{Car}}}, \bibinfo {author} {\bibfnamefont {C.}~\bibnamefont {{Cavazzoni}}}, \bibinfo {author} {\bibfnamefont {D.}~\bibnamefont {{Ceresoli}}}, \bibinfo {author} {\bibfnamefont {M.}~\bibnamefont {{Cococcioni}}}, \bibinfo {author} {\bibfnamefont {N.}~\bibnamefont {{Colonna}}}, \bibinfo {author} {\bibfnamefont {I.}~\bibnamefont {{Carnimeo}}}, \bibinfo {author} {\bibfnamefont {A.}~\bibnamefont {{Dal Corso}}}, \bibinfo {author} {\bibfnamefont {S.}~\bibnamefont {{de Gironcoli}}}, \bibinfo {author} {\bibfnamefont {P.}~\bibnamefont {{Delugas}}},
  \bibinfo {author} {\bibfnamefont {R.~A.}\ \bibnamefont {{DiStasio}}, \bibfnamefont {Jr.}}, \bibinfo {author} {\bibfnamefont {A.}~\bibnamefont {{Ferretti}}}, \bibinfo {author} {\bibfnamefont {A.}~\bibnamefont {{Floris}}}, \bibinfo {author} {\bibfnamefont {G.}~\bibnamefont {{Fratesi}}}, \bibinfo {author} {\bibfnamefont {G.}~\bibnamefont {{Fugallo}}}, \bibinfo {author} {\bibfnamefont {R.}~\bibnamefont {{Gebauer}}}, \bibinfo {author} {\bibfnamefont {U.}~\bibnamefont {{Gerstmann}}}, \bibinfo {author} {\bibfnamefont {F.}~\bibnamefont {{Giustino}}}, \bibinfo {author} {\bibfnamefont {T.}~\bibnamefont {{Gorni}}}, \bibinfo {author} {\bibfnamefont {J.}~\bibnamefont {{Jia}}}, \bibinfo {author} {\bibfnamefont {M.}~\bibnamefont {{Kawamura}}}, \bibinfo {author} {\bibfnamefont {H.~Y.}\ \bibnamefont {{Ko}}}, \bibinfo {author} {\bibfnamefont {A.}~\bibnamefont {{Kokalj}}}, \bibinfo {author} {\bibfnamefont {E.}~\bibnamefont {{K{\"u}{\c{c}}{\"u}kbenli}}}, \bibinfo {author} {\bibfnamefont {M.}~\bibnamefont {{Lazzeri}}}, \bibinfo
  {author} {\bibfnamefont {M.}~\bibnamefont {{Marsili}}}, \bibinfo {author} {\bibfnamefont {N.}~\bibnamefont {{Marzari}}}, \bibinfo {author} {\bibfnamefont {F.}~\bibnamefont {{Mauri}}}, \bibinfo {author} {\bibfnamefont {N.~L.}\ \bibnamefont {{Nguyen}}}, \bibinfo {author} {\bibfnamefont {H.~V.}\ \bibnamefont {{Nguyen}}}, \bibinfo {author} {\bibfnamefont {A.}~\bibnamefont {{Otero-de-la-Roza}}}, \bibinfo {author} {\bibfnamefont {L.}~\bibnamefont {{Paulatto}}}, \bibinfo {author} {\bibfnamefont {S.}~\bibnamefont {{Ponc{\'e}}}}, \bibinfo {author} {\bibfnamefont {D.}~\bibnamefont {{Rocca}}}, \bibinfo {author} {\bibfnamefont {R.}~\bibnamefont {{Sabatini}}}, \bibinfo {author} {\bibfnamefont {B.}~\bibnamefont {{Santra}}}, \bibinfo {author} {\bibfnamefont {M.}~\bibnamefont {{Schlipf}}}, \bibinfo {author} {\bibfnamefont {A.~P.}\ \bibnamefont {{Seitsonen}}}, \bibinfo {author} {\bibfnamefont {A.}~\bibnamefont {{Smogunov}}}, \bibinfo {author} {\bibfnamefont {I.}~\bibnamefont {{Timrov}}}, \bibinfo {author} {\bibfnamefont
  {T.}~\bibnamefont {{Thonhauser}}}, \bibinfo {author} {\bibfnamefont {P.}~\bibnamefont {{Umari}}}, \bibinfo {author} {\bibfnamefont {N.}~\bibnamefont {{Vast}}}, \bibinfo {author} {\bibfnamefont {X.}~\bibnamefont {{Wu}}},\ and\ \bibinfo {author} {\bibfnamefont {S.}~\bibnamefont {{Baroni}}},\ }\bibfield  {title} {\bibinfo {title} {{Advanced capabilities for materials modelling with Quantum ESPRESSO}},\ }\href {https://doi.org/10.1088/1361-648X/aa8f79} {\bibfield  {journal} {\bibinfo  {journal} {J. Phys. Condens. Matter}\ }\textbf {\bibinfo {volume} {29}},\ \bibinfo {pages} {465901} (\bibinfo {year} {2017})}\BibitemShut {NoStop}%
\bibitem [{\citenamefont {{Perdew}}\ \emph {et~al.}(1996)\citenamefont {{Perdew}}, \citenamefont {{Burke}},\ and\ \citenamefont {{Ernzerhof}}}]{perdew1996generalized}%
  \BibitemOpen
  \bibfield  {author} {\bibinfo {author} {\bibfnamefont {J.~P.}\ \bibnamefont {{Perdew}}}, \bibinfo {author} {\bibfnamefont {K.}~\bibnamefont {{Burke}}},\ and\ \bibinfo {author} {\bibfnamefont {M.}~\bibnamefont {{Ernzerhof}}},\ }\bibfield  {title} {\bibinfo {title} {{Generalized Gradient Approximation Made Simple}},\ }\href {https://doi.org/10.1103/PhysRevLett.77.3865} {\bibfield  {journal} {\bibinfo  {journal} {\prl}\ }\textbf {\bibinfo {volume} {77}},\ \bibinfo {pages} {3865} (\bibinfo {year} {1996})}\BibitemShut {NoStop}%
\bibitem [{\citenamefont {{Hamann}}(2013)}]{hamann2013optimized}%
  \BibitemOpen
  \bibfield  {author} {\bibinfo {author} {\bibfnamefont {D.~R.}\ \bibnamefont {{Hamann}}},\ }\bibfield  {title} {\bibinfo {title} {{Optimized norm-conserving Vanderbilt pseudopotentials}},\ }\href {https://doi.org/10.1103/PhysRevB.88.085117} {\bibfield  {journal} {\bibinfo  {journal} {\prb}\ }\textbf {\bibinfo {volume} {88}},\ \bibinfo {pages} {085117} (\bibinfo {year} {2013})}\BibitemShut {NoStop}%
\bibitem [{\citenamefont {Grimme}(2006)}]{grimme2006semiempirical}%
  \BibitemOpen
  \bibfield  {author} {\bibinfo {author} {\bibfnamefont {S.}~\bibnamefont {Grimme}},\ }\bibfield  {title} {\bibinfo {title} {{Semiempirical GGA-type density functional constructed with a long-range dispersion correction}},\ }\href {https://doi.org/10.1002/jcc.20495} {\bibfield  {journal} {\bibinfo  {journal} {J. Comput. Chem.}\ }\textbf {\bibinfo {volume} {27}},\ \bibinfo {pages} {1787} (\bibinfo {year} {2006})}\BibitemShut {NoStop}%
\bibitem [{\citenamefont {{Chadi}}(1977)}]{monkhorst1976special}%
  \BibitemOpen
  \bibfield  {author} {\bibinfo {author} {\bibfnamefont {D.~J.}\ \bibnamefont {{Chadi}}},\ }\bibfield  {title} {\bibinfo {title} {{Special points for Brillouin-zone integrations}},\ }\href {https://doi.org/10.1103/PhysRevB.16.1746} {\bibfield  {journal} {\bibinfo  {journal} {\prb}\ }\textbf {\bibinfo {volume} {16}},\ \bibinfo {pages} {1746} (\bibinfo {year} {1977})}\BibitemShut {NoStop}%
\bibitem [{\citenamefont {{Kim}}\ \emph {et~al.}(2021)\citenamefont {{Kim}}, \citenamefont {{Kim}}, \citenamefont {{Kim}}, \citenamefont {{Kwon}}, \citenamefont {{Kim}},\ and\ \citenamefont {{Kim}}}]{kim2021direct}%
  \BibitemOpen
  \bibfield  {author} {\bibinfo {author} {\bibfnamefont {K.}~\bibnamefont {{Kim}}}, \bibinfo {author} {\bibfnamefont {H.}~\bibnamefont {{Kim}}}, \bibinfo {author} {\bibfnamefont {J.}~\bibnamefont {{Kim}}}, \bibinfo {author} {\bibfnamefont {C.}~\bibnamefont {{Kwon}}}, \bibinfo {author} {\bibfnamefont {J.~S.}\ \bibnamefont {{Kim}}},\ and\ \bibinfo {author} {\bibfnamefont {B.~J.}\ \bibnamefont {{Kim}}},\ }\bibfield  {title} {\bibinfo {title} {{Direct observation of excitonic instability in Ta$_{2}$NiSe$_{5}$}},\ }\href {https://doi.org/10.1038/s41467-021-22133-z} {\bibfield  {journal} {\bibinfo  {journal} {Nat. Commun.}\ }\textbf {\bibinfo {volume} {12}},\ \bibinfo {pages} {1969} (\bibinfo {year} {2021})}\BibitemShut {NoStop}%
\bibitem [{\citenamefont {{Volkov}}\ \emph {et~al.}(2021)\citenamefont {{Volkov}}, \citenamefont {{Ye}}, \citenamefont {{Lohani}}, \citenamefont {{Feldman}}, \citenamefont {{Kanigel}},\ and\ \citenamefont {{Blumberg}}}]{volkov2021critical}%
  \BibitemOpen
  \bibfield  {author} {\bibinfo {author} {\bibfnamefont {P.~A.}\ \bibnamefont {{Volkov}}}, \bibinfo {author} {\bibfnamefont {M.}~\bibnamefont {{Ye}}}, \bibinfo {author} {\bibfnamefont {H.}~\bibnamefont {{Lohani}}}, \bibinfo {author} {\bibfnamefont {I.}~\bibnamefont {{Feldman}}}, \bibinfo {author} {\bibfnamefont {A.}~\bibnamefont {{Kanigel}}},\ and\ \bibinfo {author} {\bibfnamefont {G.}~\bibnamefont {{Blumberg}}},\ }\bibfield  {title} {\bibinfo {title} {{Critical charge fluctuations and emergent coherence in a strongly correlated excitonic insulator}},\ }\href {https://doi.org/10.1038/s41535-021-00351-4} {\bibfield  {journal} {\bibinfo  {journal} {npj Quantum Mater.}\ }\textbf {\bibinfo {volume} {6}},\ \bibinfo {pages} {52} (\bibinfo {year} {2021})}\BibitemShut {NoStop}%
\bibitem [{\citenamefont {{Watson}}\ \emph {et~al.}(2020)\citenamefont {{Watson}}, \citenamefont {{Markovi{\'c}}}, \citenamefont {{Morales}}, \citenamefont {{Le F{\`e}vre}}, \citenamefont {{Merz}}, \citenamefont {{Haghighirad}},\ and\ \citenamefont {{King}}}]{watson2020band}%
  \BibitemOpen
  \bibfield  {author} {\bibinfo {author} {\bibfnamefont {M.~D.}\ \bibnamefont {{Watson}}}, \bibinfo {author} {\bibfnamefont {I.}~\bibnamefont {{Markovi{\'c}}}}, \bibinfo {author} {\bibfnamefont {E.~A.}\ \bibnamefont {{Morales}}}, \bibinfo {author} {\bibfnamefont {P.}~\bibnamefont {{Le F{\`e}vre}}}, \bibinfo {author} {\bibfnamefont {M.}~\bibnamefont {{Merz}}}, \bibinfo {author} {\bibfnamefont {A.~A.}\ \bibnamefont {{Haghighirad}}},\ and\ \bibinfo {author} {\bibfnamefont {P.~D.~C.}\ \bibnamefont {{King}}},\ }\bibfield  {title} {\bibinfo {title} {{Band hybridization at the semimetal-semiconductor transition of Ta$_{2}$NiSe$_{5}$ enabled by mirror-symmetry breaking}},\ }\href {https://doi.org/10.1103/PhysRevResearch.2.013236} {\bibfield  {journal} {\bibinfo  {journal} {Phys. Rev. Res.}\ }\textbf {\bibinfo {volume} {2}},\ \bibinfo {pages} {013236} (\bibinfo {year} {2020})}\BibitemShut {NoStop}%
\bibitem [{\citenamefont {{Hellmann}}\ \emph {et~al.}(2012)\citenamefont {{Hellmann}}, \citenamefont {{Rohwer}}, \citenamefont {{Kall{\"a}ne}}, \citenamefont {{Hanff}}, \citenamefont {{Sohrt}}, \citenamefont {{Stange}}, \citenamefont {{Carr}}, \citenamefont {{Murnane}}, \citenamefont {{Kapteyn}}, \citenamefont {{Kipp}}, \citenamefont {{Bauer}},\ and\ \citenamefont {{Rossnagel}}}]{hellmann2012time}%
  \BibitemOpen
  \bibfield  {author} {\bibinfo {author} {\bibfnamefont {S.}~\bibnamefont {{Hellmann}}}, \bibinfo {author} {\bibfnamefont {T.}~\bibnamefont {{Rohwer}}}, \bibinfo {author} {\bibfnamefont {M.}~\bibnamefont {{Kall{\"a}ne}}}, \bibinfo {author} {\bibfnamefont {K.}~\bibnamefont {{Hanff}}}, \bibinfo {author} {\bibfnamefont {C.}~\bibnamefont {{Sohrt}}}, \bibinfo {author} {\bibfnamefont {A.}~\bibnamefont {{Stange}}}, \bibinfo {author} {\bibfnamefont {A.}~\bibnamefont {{Carr}}}, \bibinfo {author} {\bibfnamefont {M.~M.}\ \bibnamefont {{Murnane}}}, \bibinfo {author} {\bibfnamefont {H.~C.}\ \bibnamefont {{Kapteyn}}}, \bibinfo {author} {\bibfnamefont {L.}~\bibnamefont {{Kipp}}}, \bibinfo {author} {\bibfnamefont {M.}~\bibnamefont {{Bauer}}},\ and\ \bibinfo {author} {\bibfnamefont {K.}~\bibnamefont {{Rossnagel}}},\ }\bibfield  {title} {\bibinfo {title} {{Time-domain classification of charge-density-wave insulators}},\ }\href {https://doi.org/10.1038/ncomms2078} {\bibfield  {journal} {\bibinfo  {journal} {Nat. Commun.}\
  }\textbf {\bibinfo {volume} {3}},\ \bibinfo {pages} {1069} (\bibinfo {year} {2012})}\BibitemShut {NoStop}%
\bibitem [{\citenamefont {{Mor}}\ \emph {et~al.}(2017)\citenamefont {{Mor}}, \citenamefont {{Herzog}}, \citenamefont {{Gole{\v{z}}}}, \citenamefont {{Werner}}, \citenamefont {{Eckstein}}, \citenamefont {{Katayama}}, \citenamefont {{Nohara}}, \citenamefont {{Takagi}}, \citenamefont {{Mizokawa}}, \citenamefont {{Monney}},\ and\ \citenamefont {{St{\"a}hler}}}]{mor2017ultrafast}%
  \BibitemOpen
  \bibfield  {author} {\bibinfo {author} {\bibfnamefont {S.}~\bibnamefont {{Mor}}}, \bibinfo {author} {\bibfnamefont {M.}~\bibnamefont {{Herzog}}}, \bibinfo {author} {\bibfnamefont {D.}~\bibnamefont {{Gole{\v{z}}}}}, \bibinfo {author} {\bibfnamefont {P.}~\bibnamefont {{Werner}}}, \bibinfo {author} {\bibfnamefont {M.}~\bibnamefont {{Eckstein}}}, \bibinfo {author} {\bibfnamefont {N.}~\bibnamefont {{Katayama}}}, \bibinfo {author} {\bibfnamefont {M.}~\bibnamefont {{Nohara}}}, \bibinfo {author} {\bibfnamefont {H.}~\bibnamefont {{Takagi}}}, \bibinfo {author} {\bibfnamefont {T.}~\bibnamefont {{Mizokawa}}}, \bibinfo {author} {\bibfnamefont {C.}~\bibnamefont {{Monney}}},\ and\ \bibinfo {author} {\bibfnamefont {J.}~\bibnamefont {{St{\"a}hler}}},\ }\bibfield  {title} {\bibinfo {title} {{Ultrafast Electronic Band Gap Control in an Excitonic Insulator}},\ }\href {https://doi.org/10.1103/PhysRevLett.119.086401} {\bibfield  {journal} {\bibinfo  {journal} {\prl}\ }\textbf {\bibinfo {volume} {119}},\ \bibinfo {pages} {086401}
  (\bibinfo {year} {2017})}\BibitemShut {NoStop}%
\bibitem [{\citenamefont {{Okazaki}}\ \emph {et~al.}(2018)\citenamefont {{Okazaki}}, \citenamefont {{Ogawa}}, \citenamefont {{Suzuki}}, \citenamefont {{Yamamoto}}, \citenamefont {{Someya}}, \citenamefont {{Michimae}}, \citenamefont {{Watanabe}}, \citenamefont {{Lu}}, \citenamefont {{Nohara}}, \citenamefont {{Takagi}}, \citenamefont {{Katayama}}, \citenamefont {{Sawa}}, \citenamefont {{Fujisawa}}, \citenamefont {{Kanai}}, \citenamefont {{Ishii}}, \citenamefont {{Itatani}}, \citenamefont {{Mizokawa}},\ and\ \citenamefont {{Shin}}}]{okazaki2018photo}%
  \BibitemOpen
  \bibfield  {author} {\bibinfo {author} {\bibfnamefont {K.}~\bibnamefont {{Okazaki}}}, \bibinfo {author} {\bibfnamefont {Y.}~\bibnamefont {{Ogawa}}}, \bibinfo {author} {\bibfnamefont {T.}~\bibnamefont {{Suzuki}}}, \bibinfo {author} {\bibfnamefont {T.}~\bibnamefont {{Yamamoto}}}, \bibinfo {author} {\bibfnamefont {T.}~\bibnamefont {{Someya}}}, \bibinfo {author} {\bibfnamefont {S.}~\bibnamefont {{Michimae}}}, \bibinfo {author} {\bibfnamefont {M.}~\bibnamefont {{Watanabe}}}, \bibinfo {author} {\bibfnamefont {Y.}~\bibnamefont {{Lu}}}, \bibinfo {author} {\bibfnamefont {M.}~\bibnamefont {{Nohara}}}, \bibinfo {author} {\bibfnamefont {H.}~\bibnamefont {{Takagi}}}, \bibinfo {author} {\bibfnamefont {N.}~\bibnamefont {{Katayama}}}, \bibinfo {author} {\bibfnamefont {H.}~\bibnamefont {{Sawa}}}, \bibinfo {author} {\bibfnamefont {M.}~\bibnamefont {{Fujisawa}}}, \bibinfo {author} {\bibfnamefont {T.}~\bibnamefont {{Kanai}}}, \bibinfo {author} {\bibfnamefont {N.}~\bibnamefont {{Ishii}}}, \bibinfo {author} {\bibfnamefont
  {J.}~\bibnamefont {{Itatani}}}, \bibinfo {author} {\bibfnamefont {T.}~\bibnamefont {{Mizokawa}}},\ and\ \bibinfo {author} {\bibfnamefont {S.}~\bibnamefont {{Shin}}},\ }\bibfield  {title} {\bibinfo {title} {{Photo-induced semimetallic states realised in electron-hole coupled insulators}},\ }\href {https://doi.org/10.1038/s41467-018-06801-1} {\bibfield  {journal} {\bibinfo  {journal} {Nat. Commun.}\ }\textbf {\bibinfo {volume} {9}},\ \bibinfo {pages} {4322} (\bibinfo {year} {2018})}\BibitemShut {NoStop}%
\bibitem [{\citenamefont {{Tang}}\ \emph {et~al.}(2020)\citenamefont {{Tang}}, \citenamefont {{Wang}}, \citenamefont {{Duan}}, \citenamefont {{Yang}}, \citenamefont {{Huang}}, \citenamefont {{Guo}}, \citenamefont {{Qian}},\ and\ \citenamefont {{Zhang}}}]{tang2020non}%
  \BibitemOpen
  \bibfield  {author} {\bibinfo {author} {\bibfnamefont {T.}~\bibnamefont {{Tang}}}, \bibinfo {author} {\bibfnamefont {H.}~\bibnamefont {{Wang}}}, \bibinfo {author} {\bibfnamefont {S.}~\bibnamefont {{Duan}}}, \bibinfo {author} {\bibfnamefont {Y.}~\bibnamefont {{Yang}}}, \bibinfo {author} {\bibfnamefont {C.}~\bibnamefont {{Huang}}}, \bibinfo {author} {\bibfnamefont {Y.}~\bibnamefont {{Guo}}}, \bibinfo {author} {\bibfnamefont {D.}~\bibnamefont {{Qian}}},\ and\ \bibinfo {author} {\bibfnamefont {W.}~\bibnamefont {{Zhang}}},\ }\bibfield  {title} {\bibinfo {title} {{Non-Coulomb strong electron-hole binding in Ta$_{2}$NiSe$_{5}$ revealed by time- and angle-resolved photoemission spectroscopy}},\ }\href {https://doi.org/10.1103/PhysRevB.101.235148} {\bibfield  {journal} {\bibinfo  {journal} {\prb}\ }\textbf {\bibinfo {volume} {101}},\ \bibinfo {pages} {235148} (\bibinfo {year} {2020})}\BibitemShut {NoStop}%
\bibitem [{\citenamefont {{Suzuki}}\ \emph {et~al.}(2021)\citenamefont {{Suzuki}}, \citenamefont {{Shinohara}}, \citenamefont {{Lu}}, \citenamefont {{Watanabe}}, \citenamefont {{Xu}}, \citenamefont {{Ishikawa}}, \citenamefont {{Takagi}}, \citenamefont {{Nohara}}, \citenamefont {{Katayama}}, \citenamefont {{Sawa}}, \citenamefont {{Fujisawa}}, \citenamefont {{Kanai}}, \citenamefont {{Itatani}}, \citenamefont {{Mizokawa}}, \citenamefont {{Shin}},\ and\ \citenamefont {{Okazaki}}}]{suzuki2021detecting}%
  \BibitemOpen
  \bibfield  {author} {\bibinfo {author} {\bibfnamefont {T.}~\bibnamefont {{Suzuki}}}, \bibinfo {author} {\bibfnamefont {Y.}~\bibnamefont {{Shinohara}}}, \bibinfo {author} {\bibfnamefont {Y.}~\bibnamefont {{Lu}}}, \bibinfo {author} {\bibfnamefont {M.}~\bibnamefont {{Watanabe}}}, \bibinfo {author} {\bibfnamefont {J.}~\bibnamefont {{Xu}}}, \bibinfo {author} {\bibfnamefont {K.~L.}\ \bibnamefont {{Ishikawa}}}, \bibinfo {author} {\bibfnamefont {H.}~\bibnamefont {{Takagi}}}, \bibinfo {author} {\bibfnamefont {M.}~\bibnamefont {{Nohara}}}, \bibinfo {author} {\bibfnamefont {N.}~\bibnamefont {{Katayama}}}, \bibinfo {author} {\bibfnamefont {H.}~\bibnamefont {{Sawa}}}, \bibinfo {author} {\bibfnamefont {M.}~\bibnamefont {{Fujisawa}}}, \bibinfo {author} {\bibfnamefont {T.}~\bibnamefont {{Kanai}}}, \bibinfo {author} {\bibfnamefont {J.}~\bibnamefont {{Itatani}}}, \bibinfo {author} {\bibfnamefont {T.}~\bibnamefont {{Mizokawa}}}, \bibinfo {author} {\bibfnamefont {S.}~\bibnamefont {{Shin}}},\ and\ \bibinfo {author}
  {\bibfnamefont {K.}~\bibnamefont {{Okazaki}}},\ }\bibfield  {title} {\bibinfo {title} {{Detecting electron-phonon coupling during photoinduced phase transition}},\ }\href {https://doi.org/10.1103/PhysRevB.103.L121105} {\bibfield  {journal} {\bibinfo  {journal} {\prb}\ }\textbf {\bibinfo {volume} {103}},\ \bibinfo {pages} {L121105} (\bibinfo {year} {2021})}\BibitemShut {NoStop}%
\bibitem [{\citenamefont {{Saha}}\ \emph {et~al.}(2021)\citenamefont {{Saha}}, \citenamefont {{Gole{\v{z}}}}, \citenamefont {{De Ninno}}, \citenamefont {{Mravlje}}, \citenamefont {{Murakami}}, \citenamefont {{Ressel}}, \citenamefont {{Stupar}},\ and\ \citenamefont {{Ribi{\v{c}}}}}]{saha2021photoinduced}%
  \BibitemOpen
  \bibfield  {author} {\bibinfo {author} {\bibfnamefont {T.}~\bibnamefont {{Saha}}}, \bibinfo {author} {\bibfnamefont {D.}~\bibnamefont {{Gole{\v{z}}}}}, \bibinfo {author} {\bibfnamefont {G.}~\bibnamefont {{De Ninno}}}, \bibinfo {author} {\bibfnamefont {J.}~\bibnamefont {{Mravlje}}}, \bibinfo {author} {\bibfnamefont {Y.}~\bibnamefont {{Murakami}}}, \bibinfo {author} {\bibfnamefont {B.}~\bibnamefont {{Ressel}}}, \bibinfo {author} {\bibfnamefont {M.}~\bibnamefont {{Stupar}}},\ and\ \bibinfo {author} {\bibfnamefont {P.~R.}\ \bibnamefont {{Ribi{\v{c}}}}},\ }\bibfield  {title} {\bibinfo {title} {{Photoinduced phase transition and associated timescales in the excitonic insulator Ta$_{2}$NiSe$_{5}$}},\ }\href {https://doi.org/10.1103/PhysRevB.103.144304} {\bibfield  {journal} {\bibinfo  {journal} {\prb}\ }\textbf {\bibinfo {volume} {103}},\ \bibinfo {pages} {144304} (\bibinfo {year} {2021})}\BibitemShut {NoStop}%
\bibitem [{\citenamefont {{Baldini}}\ \emph {et~al.}(2023)\citenamefont {{Baldini}}, \citenamefont {{Zong}}, \citenamefont {{Choi}}, \citenamefont {{Lee}}, \citenamefont {{Michael}}, \citenamefont {{Windgaetter}}, \citenamefont {{Mazin}}, \citenamefont {{Latini}}, \citenamefont {{Azoury}}, \citenamefont {{Lv}}, \citenamefont {{Kogar}}, \citenamefont {{Su}}, \citenamefont {{Wang}}, \citenamefont {{Lu}}, \citenamefont {{Takayama}}, \citenamefont {{Takagi}}, \citenamefont {{Millis}}, \citenamefont {{Rubio}}, \citenamefont {{Demler}},\ and\ \citenamefont {{Gedik}}}]{baldini2023spontaneous}%
  \BibitemOpen
  \bibfield  {author} {\bibinfo {author} {\bibfnamefont {E.}~\bibnamefont {{Baldini}}}, \bibinfo {author} {\bibfnamefont {A.}~\bibnamefont {{Zong}}}, \bibinfo {author} {\bibfnamefont {D.}~\bibnamefont {{Choi}}}, \bibinfo {author} {\bibfnamefont {C.}~\bibnamefont {{Lee}}}, \bibinfo {author} {\bibfnamefont {M.~H.}\ \bibnamefont {{Michael}}}, \bibinfo {author} {\bibfnamefont {L.}~\bibnamefont {{Windgaetter}}}, \bibinfo {author} {\bibfnamefont {I.~I.}\ \bibnamefont {{Mazin}}}, \bibinfo {author} {\bibfnamefont {S.}~\bibnamefont {{Latini}}}, \bibinfo {author} {\bibfnamefont {D.}~\bibnamefont {{Azoury}}}, \bibinfo {author} {\bibfnamefont {B.}~\bibnamefont {{Lv}}}, \bibinfo {author} {\bibfnamefont {A.}~\bibnamefont {{Kogar}}}, \bibinfo {author} {\bibfnamefont {Y.}~\bibnamefont {{Su}}}, \bibinfo {author} {\bibfnamefont {Y.}~\bibnamefont {{Wang}}}, \bibinfo {author} {\bibfnamefont {Y.}~\bibnamefont {{Lu}}}, \bibinfo {author} {\bibfnamefont {T.}~\bibnamefont {{Takayama}}}, \bibinfo {author} {\bibfnamefont
  {H.}~\bibnamefont {{Takagi}}}, \bibinfo {author} {\bibfnamefont {A.~J.}\ \bibnamefont {{Millis}}}, \bibinfo {author} {\bibfnamefont {A.}~\bibnamefont {{Rubio}}}, \bibinfo {author} {\bibfnamefont {E.}~\bibnamefont {{Demler}}},\ and\ \bibinfo {author} {\bibfnamefont {N.}~\bibnamefont {{Gedik}}},\ }\bibfield  {title} {\bibinfo {title} {{The spontaneous symmetry breaking in Ta$_{2}$NiSe$_{5}$ is structural in nature}},\ }\href {https://doi.org/10.1073/pnas.2221688120} {\bibfield  {journal} {\bibinfo  {journal} {Proc. Natl. Acad. Sci. U.S.A.}\ }\textbf {\bibinfo {volume} {120}},\ \bibinfo {pages} {e2221688120} (\bibinfo {year} {2023})}\BibitemShut {NoStop}%
\bibitem [{\citenamefont {{Bretscher}}\ \emph {et~al.}(2021)\citenamefont {{Bretscher}}, \citenamefont {{Andrich}}, \citenamefont {{Telang}}, \citenamefont {{Singh}}, \citenamefont {{Harnagea}}, \citenamefont {{Sood}},\ and\ \citenamefont {{Rao}}}]{bretscher2021ultrafast}%
  \BibitemOpen
  \bibfield  {author} {\bibinfo {author} {\bibfnamefont {H.~M.}\ \bibnamefont {{Bretscher}}}, \bibinfo {author} {\bibfnamefont {P.}~\bibnamefont {{Andrich}}}, \bibinfo {author} {\bibfnamefont {P.}~\bibnamefont {{Telang}}}, \bibinfo {author} {\bibfnamefont {A.}~\bibnamefont {{Singh}}}, \bibinfo {author} {\bibfnamefont {L.}~\bibnamefont {{Harnagea}}}, \bibinfo {author} {\bibfnamefont {A.~K.}\ \bibnamefont {{Sood}}},\ and\ \bibinfo {author} {\bibfnamefont {A.}~\bibnamefont {{Rao}}},\ }\bibfield  {title} {\bibinfo {title} {{Ultrafast melting and recovery of collective order in the excitonic insulator Ta$_{2}$NiSe$_{5}$}},\ }\href {https://doi.org/10.1038/s41467-021-21929-3} {\bibfield  {journal} {\bibinfo  {journal} {Nat. Commun.}\ }\textbf {\bibinfo {volume} {12}},\ \bibinfo {pages} {1699} (\bibinfo {year} {2021})}\BibitemShut {NoStop}%
\bibitem [{\citenamefont {{Miyamoto}}\ \emph {et~al.}(2022)\citenamefont {{Miyamoto}}, \citenamefont {{Mizui}}, \citenamefont {{Takamura}}, \citenamefont {{Hirata}}, \citenamefont {{Yamakawa}}, \citenamefont {{Morimoto}}, \citenamefont {{Terashige}}, \citenamefont {{Kida}}, \citenamefont {{Nakano}}, \citenamefont {{Sawa}},\ and\ \citenamefont {{Okamoto}}}]{miyamoto2022charge}%
  \BibitemOpen
  \bibfield  {author} {\bibinfo {author} {\bibfnamefont {T.}~\bibnamefont {{Miyamoto}}}, \bibinfo {author} {\bibfnamefont {M.}~\bibnamefont {{Mizui}}}, \bibinfo {author} {\bibfnamefont {N.}~\bibnamefont {{Takamura}}}, \bibinfo {author} {\bibfnamefont {J.}~\bibnamefont {{Hirata}}}, \bibinfo {author} {\bibfnamefont {H.}~\bibnamefont {{Yamakawa}}}, \bibinfo {author} {\bibfnamefont {T.}~\bibnamefont {{Morimoto}}}, \bibinfo {author} {\bibfnamefont {T.}~\bibnamefont {{Terashige}}}, \bibinfo {author} {\bibfnamefont {N.}~\bibnamefont {{Kida}}}, \bibinfo {author} {\bibfnamefont {A.}~\bibnamefont {{Nakano}}}, \bibinfo {author} {\bibfnamefont {H.}~\bibnamefont {{Sawa}}},\ and\ \bibinfo {author} {\bibfnamefont {H.}~\bibnamefont {{Okamoto}}},\ }\bibfield  {title} {\bibinfo {title} {{Charge and Lattice Dynamics in Excitonic Insulator Ta$_{2}$NiSe$_{5}$ Investigated Using Ultrafast Reflection Spectroscopy}},\ }\href {https://doi.org/10.7566/JPSJ.91.023701} {\bibfield  {journal} {\bibinfo  {journal} {J. Phys. Soc. Japan}\
  }\textbf {\bibinfo {volume} {91}},\ \bibinfo {pages} {023701} (\bibinfo {year} {2022})}\BibitemShut {NoStop}%
\bibitem [{\citenamefont {{Katsumi}}\ \emph {et~al.}(2023)\citenamefont {{Katsumi}}, \citenamefont {{Alekhin}}, \citenamefont {{Souliou}}, \citenamefont {{Merz}}, \citenamefont {{Haghighirad}}, \citenamefont {{Le Tacon}}, \citenamefont {{Houver}}, \citenamefont {{Cazayous}}, \citenamefont {{Sacuto}},\ and\ \citenamefont {{Gallais}}}]{katsumi2023disentangling}%
  \BibitemOpen
  \bibfield  {author} {\bibinfo {author} {\bibfnamefont {K.}~\bibnamefont {{Katsumi}}}, \bibinfo {author} {\bibfnamefont {A.}~\bibnamefont {{Alekhin}}}, \bibinfo {author} {\bibfnamefont {S.-M.}\ \bibnamefont {{Souliou}}}, \bibinfo {author} {\bibfnamefont {M.}~\bibnamefont {{Merz}}}, \bibinfo {author} {\bibfnamefont {A.-A.}\ \bibnamefont {{Haghighirad}}}, \bibinfo {author} {\bibfnamefont {M.}~\bibnamefont {{Le Tacon}}}, \bibinfo {author} {\bibfnamefont {S.}~\bibnamefont {{Houver}}}, \bibinfo {author} {\bibfnamefont {M.}~\bibnamefont {{Cazayous}}}, \bibinfo {author} {\bibfnamefont {A.}~\bibnamefont {{Sacuto}}},\ and\ \bibinfo {author} {\bibfnamefont {Y.}~\bibnamefont {{Gallais}}},\ }\bibfield  {title} {\bibinfo {title} {{Disentangling Lattice and Electronic Instabilities in the Excitonic Insulator Candidate Ta$_{2}$NiSe$_{5}$ by Nonequilibrium Spectroscopy}},\ }\href {https://doi.org/10.1103/PhysRevLett.130.106904} {\bibfield  {journal} {\bibinfo  {journal} {\prl}\ }\textbf {\bibinfo {volume} {130}},\ \bibinfo
  {pages} {106904} (\bibinfo {year} {2023})}\BibitemShut {NoStop}%
\bibitem [{\citenamefont {{Werdehausen}}\ \emph {et~al.}(2018{\natexlab{a}})\citenamefont {{Werdehausen}}, \citenamefont {{Takayama}}, \citenamefont {{H{\"o}ppner}}, \citenamefont {{Albrecht}}, \citenamefont {{Rost}}, \citenamefont {{Lu}}, \citenamefont {{Manske}}, \citenamefont {{Takagi}},\ and\ \citenamefont {{Kaiser}}}]{werdehausen2018coherent}%
  \BibitemOpen
  \bibfield  {author} {\bibinfo {author} {\bibfnamefont {D.}~\bibnamefont {{Werdehausen}}}, \bibinfo {author} {\bibfnamefont {T.}~\bibnamefont {{Takayama}}}, \bibinfo {author} {\bibfnamefont {M.}~\bibnamefont {{H{\"o}ppner}}}, \bibinfo {author} {\bibfnamefont {G.}~\bibnamefont {{Albrecht}}}, \bibinfo {author} {\bibfnamefont {A.~W.}\ \bibnamefont {{Rost}}}, \bibinfo {author} {\bibfnamefont {Y.}~\bibnamefont {{Lu}}}, \bibinfo {author} {\bibfnamefont {D.}~\bibnamefont {{Manske}}}, \bibinfo {author} {\bibfnamefont {H.}~\bibnamefont {{Takagi}}},\ and\ \bibinfo {author} {\bibfnamefont {S.}~\bibnamefont {{Kaiser}}},\ }\bibfield  {title} {\bibinfo {title} {{Coherent order parameter oscillations in the ground state of the excitonic insulator Ta$_2$NiSe$_5$}},\ }\href {https://doi.org/10.1126/sciadv.aap8652} {\bibfield  {journal} {\bibinfo  {journal} {Sci. Adv.}\ }\textbf {\bibinfo {volume} {4}},\ \bibinfo {pages} {eaap8652} (\bibinfo {year} {2018}{\natexlab{a}})}\BibitemShut {NoStop}%
\bibitem [{\citenamefont {{Mor}}\ \emph {et~al.}(2022)\citenamefont {{Mor}}, \citenamefont {{Herzog}}, \citenamefont {{Monney}},\ and\ \citenamefont {{St{\"a}hler}}}]{Mor2022ultrafast}%
  \BibitemOpen
  \bibfield  {author} {\bibinfo {author} {\bibfnamefont {S.}~\bibnamefont {{Mor}}}, \bibinfo {author} {\bibfnamefont {M.}~\bibnamefont {{Herzog}}}, \bibinfo {author} {\bibfnamefont {C.}~\bibnamefont {{Monney}}},\ and\ \bibinfo {author} {\bibfnamefont {J.}~\bibnamefont {{St{\"a}hler}}},\ }\bibfield  {title} {\bibinfo {title} {{Ultrafast charge carrier and exciton dynamics in an excitonic insulator probed by time-resolved photoemission spectroscopy}},\ }\href {https://doi.org/10.1016/j.progsurf.2022.100679} {\bibfield  {journal} {\bibinfo  {journal} {Prog. Surf. Sci.}\ }\textbf {\bibinfo {volume} {97}},\ \bibinfo {pages} {100679} (\bibinfo {year} {2022})}\BibitemShut {NoStop}%
\bibitem [{\citenamefont {{Gole{\v{z}}}}\ \emph {et~al.}(2022)\citenamefont {{Gole{\v{z}}}}, \citenamefont {{Dufresne}}, \citenamefont {{Kim}}, \citenamefont {{Boschini}}, \citenamefont {{Chu}}, \citenamefont {{Murakami}}, \citenamefont {{Levy}}, \citenamefont {{Mills}}, \citenamefont {{Zhdanovich}}, \citenamefont {{Isobe}}, \citenamefont {{Takagi}}, \citenamefont {{Kaiser}}, \citenamefont {{Werner}}, \citenamefont {{Jones}}, \citenamefont {{Georges}}, \citenamefont {{Damascelli}},\ and\ \citenamefont {{Millis}}}]{golevz2022unveiling}%
  \BibitemOpen
  \bibfield  {author} {\bibinfo {author} {\bibfnamefont {D.}~\bibnamefont {{Gole{\v{z}}}}}, \bibinfo {author} {\bibfnamefont {S.~K.~Y.}\ \bibnamefont {{Dufresne}}}, \bibinfo {author} {\bibfnamefont {M.-J.}\ \bibnamefont {{Kim}}}, \bibinfo {author} {\bibfnamefont {F.}~\bibnamefont {{Boschini}}}, \bibinfo {author} {\bibfnamefont {H.}~\bibnamefont {{Chu}}}, \bibinfo {author} {\bibfnamefont {Y.}~\bibnamefont {{Murakami}}}, \bibinfo {author} {\bibfnamefont {G.}~\bibnamefont {{Levy}}}, \bibinfo {author} {\bibfnamefont {A.~K.}\ \bibnamefont {{Mills}}}, \bibinfo {author} {\bibfnamefont {S.}~\bibnamefont {{Zhdanovich}}}, \bibinfo {author} {\bibfnamefont {M.}~\bibnamefont {{Isobe}}}, \bibinfo {author} {\bibfnamefont {H.}~\bibnamefont {{Takagi}}}, \bibinfo {author} {\bibfnamefont {S.}~\bibnamefont {{Kaiser}}}, \bibinfo {author} {\bibfnamefont {P.}~\bibnamefont {{Werner}}}, \bibinfo {author} {\bibfnamefont {D.~J.}\ \bibnamefont {{Jones}}}, \bibinfo {author} {\bibfnamefont {A.}~\bibnamefont {{Georges}}}, \bibinfo {author}
  {\bibfnamefont {A.}~\bibnamefont {{Damascelli}}},\ and\ \bibinfo {author} {\bibfnamefont {A.~J.}\ \bibnamefont {{Millis}}},\ }\bibfield  {title} {\bibinfo {title} {{Unveiling the underlying interactions in Ta$_{2}$NiSe$_{5}$ from photoinduced lifetime change}},\ }\href {https://doi.org/10.1103/PhysRevB.106.L121106} {\bibfield  {journal} {\bibinfo  {journal} {\prb}\ }\textbf {\bibinfo {volume} {106}},\ \bibinfo {pages} {L121106} (\bibinfo {year} {2022})}\BibitemShut {NoStop}%
\bibitem [{\citenamefont {{Qi}}\ \emph {et~al.}(2020)\citenamefont {{Qi}}, \citenamefont {{Ma}}, \citenamefont {{Zhao}}, \citenamefont {{Cheng}}, \citenamefont {{Jiang}}, \citenamefont {{Lu}}, \citenamefont {{Jiang}}, \citenamefont {{Qian}}, \citenamefont {{Wang}}, \citenamefont {{Zhang}}, \citenamefont {{Zhu}}, \citenamefont {{Zou}}, \citenamefont {{Wan}}, \citenamefont {{Xiang}},\ and\ \citenamefont {{Zhang}}}]{qi2020breaking}%
  \BibitemOpen
  \bibfield  {author} {\bibinfo {author} {\bibfnamefont {F.}~\bibnamefont {{Qi}}}, \bibinfo {author} {\bibfnamefont {Z.}~\bibnamefont {{Ma}}}, \bibinfo {author} {\bibfnamefont {L.}~\bibnamefont {{Zhao}}}, \bibinfo {author} {\bibfnamefont {Y.}~\bibnamefont {{Cheng}}}, \bibinfo {author} {\bibfnamefont {W.}~\bibnamefont {{Jiang}}}, \bibinfo {author} {\bibfnamefont {C.}~\bibnamefont {{Lu}}}, \bibinfo {author} {\bibfnamefont {T.}~\bibnamefont {{Jiang}}}, \bibinfo {author} {\bibfnamefont {D.}~\bibnamefont {{Qian}}}, \bibinfo {author} {\bibfnamefont {Z.}~\bibnamefont {{Wang}}}, \bibinfo {author} {\bibfnamefont {W.}~\bibnamefont {{Zhang}}}, \bibinfo {author} {\bibfnamefont {P.}~\bibnamefont {{Zhu}}}, \bibinfo {author} {\bibfnamefont {X.}~\bibnamefont {{Zou}}}, \bibinfo {author} {\bibfnamefont {W.}~\bibnamefont {{Wan}}}, \bibinfo {author} {\bibfnamefont {D.}~\bibnamefont {{Xiang}}},\ and\ \bibinfo {author} {\bibfnamefont {J.}~\bibnamefont {{Zhang}}},\ }\bibfield  {title} {\bibinfo {title} {{Breaking 50 Femtosecond
  Resolution Barrier in MeV Ultrafast Electron Diffraction with a Double Bend Achromat Compressor}},\ }\href {https://doi.org/10.1103/PhysRevLett.124.134803} {\bibfield  {journal} {\bibinfo  {journal} {\prl}\ }\textbf {\bibinfo {volume} {124}},\ \bibinfo {pages} {134803} (\bibinfo {year} {2020})}\BibitemShut {NoStop}%
\bibitem [{\citenamefont {{Subedi}}(2020)}]{subedi2020orthorhombic}%
  \BibitemOpen
  \bibfield  {author} {\bibinfo {author} {\bibfnamefont {A.}~\bibnamefont {{Subedi}}},\ }\bibfield  {title} {\bibinfo {title} {{Orthorhombic-to-monoclinic transition in Ta$_{2}$NiSe$_{5}$ due to a zone-center optical phonon instability}},\ }\href {https://doi.org/10.1103/PhysRevMaterials.4.083601} {\bibfield  {journal} {\bibinfo  {journal} {Phys. Rev. Mater.}\ }\textbf {\bibinfo {volume} {4}},\ \bibinfo {pages} {083601} (\bibinfo {year} {2020})}\BibitemShut {NoStop}%
\bibitem [{\citenamefont {{Liu}}\ \emph {et~al.}(2021)\citenamefont {{Liu}}, \citenamefont {{Wu}}, \citenamefont {{Li}}, \citenamefont {{Shi}}, \citenamefont {{Wang}}, \citenamefont {{Zhang}}, \citenamefont {{Lin}}, \citenamefont {{Hu}}, \citenamefont {{Tian}}, \citenamefont {{Li}}, \citenamefont {{Dong}},\ and\ \citenamefont {{Wang}}}]{liu2021photoinduced}%
  \BibitemOpen
  \bibfield  {author} {\bibinfo {author} {\bibfnamefont {Q.~M.}\ \bibnamefont {{Liu}}}, \bibinfo {author} {\bibfnamefont {D.}~\bibnamefont {{Wu}}}, \bibinfo {author} {\bibfnamefont {Z.~A.}\ \bibnamefont {{Li}}}, \bibinfo {author} {\bibfnamefont {L.~Y.}\ \bibnamefont {{Shi}}}, \bibinfo {author} {\bibfnamefont {Z.~X.}\ \bibnamefont {{Wang}}}, \bibinfo {author} {\bibfnamefont {S.~J.}\ \bibnamefont {{Zhang}}}, \bibinfo {author} {\bibfnamefont {T.}~\bibnamefont {{Lin}}}, \bibinfo {author} {\bibfnamefont {T.~C.}\ \bibnamefont {{Hu}}}, \bibinfo {author} {\bibfnamefont {H.~F.}\ \bibnamefont {{Tian}}}, \bibinfo {author} {\bibfnamefont {J.~Q.}\ \bibnamefont {{Li}}}, \bibinfo {author} {\bibfnamefont {T.}~\bibnamefont {{Dong}}},\ and\ \bibinfo {author} {\bibfnamefont {N.~L.}\ \bibnamefont {{Wang}}},\ }\bibfield  {title} {\bibinfo {title} {{Photoinduced multistage phase transitions in Ta$_{2}$NiSe$_{5}$}},\ }\href {https://doi.org/10.1038/s41467-021-22345-3} {\bibfield  {journal} {\bibinfo  {journal} {Nat. Commun.}\
  }\textbf {\bibinfo {volume} {12}},\ \bibinfo {pages} {2050} (\bibinfo {year} {2021})}\BibitemShut {NoStop}%
\bibitem [{\citenamefont {{Gole{\v{z}}}}\ \emph {et~al.}(2020)\citenamefont {{Gole{\v{z}}}}, \citenamefont {{Sun}}, \citenamefont {{Murakami}}, \citenamefont {{Georges}},\ and\ \citenamefont {{Millis}}}]{golevz2020nonlinear}%
  \BibitemOpen
  \bibfield  {author} {\bibinfo {author} {\bibfnamefont {D.}~\bibnamefont {{Gole{\v{z}}}}}, \bibinfo {author} {\bibfnamefont {Z.}~\bibnamefont {{Sun}}}, \bibinfo {author} {\bibfnamefont {Y.}~\bibnamefont {{Murakami}}}, \bibinfo {author} {\bibfnamefont {A.}~\bibnamefont {{Georges}}},\ and\ \bibinfo {author} {\bibfnamefont {A.~J.}\ \bibnamefont {{Millis}}},\ }\bibfield  {title} {\bibinfo {title} {{Nonlinear Spectroscopy of Collective Modes in an Excitonic Insulator}},\ }\href {https://doi.org/10.1103/PhysRevLett.125.257601} {\bibfield  {journal} {\bibinfo  {journal} {\prl}\ }\textbf {\bibinfo {volume} {125}},\ \bibinfo {pages} {257601} (\bibinfo {year} {2020})}\BibitemShut {NoStop}%
\bibitem [{\citenamefont {{Guan}}\ \emph {et~al.}(2023)\citenamefont {{Guan}}, \citenamefont {{Chen}}, \citenamefont {{Chen}}, \citenamefont {{Yao}},\ and\ \citenamefont {{Meng}}}]{guan2023coherent}%
  \BibitemOpen
  \bibfield  {author} {\bibinfo {author} {\bibfnamefont {M.}~\bibnamefont {{Guan}}}, \bibinfo {author} {\bibfnamefont {D.}~\bibnamefont {{Chen}}}, \bibinfo {author} {\bibfnamefont {Q.}~\bibnamefont {{Chen}}}, \bibinfo {author} {\bibfnamefont {Y.}~\bibnamefont {{Yao}}},\ and\ \bibinfo {author} {\bibfnamefont {S.}~\bibnamefont {{Meng}}},\ }\bibfield  {title} {\bibinfo {title} {{Coherent Phonon Assisted Ultrafast Order-Parameter Reversal and Hidden Metallic State in Ta$_{2}$NiSe$_{5}$}},\ }\href {https://doi.org/10.1103/PhysRevLett.131.256503} {\bibfield  {journal} {\bibinfo  {journal} {\prl}\ }\textbf {\bibinfo {volume} {131}},\ \bibinfo {pages} {256503} (\bibinfo {year} {2023})}\BibitemShut {NoStop}%
\bibitem [{\citenamefont {{Mor}}\ \emph {et~al.}(2018)\citenamefont {{Mor}}, \citenamefont {{Herzog}}, \citenamefont {{Noack}}, \citenamefont {{Katayama}}, \citenamefont {{Nohara}}, \citenamefont {{Takagi}}, \citenamefont {{Trunschke}}, \citenamefont {{Mizokawa}}, \citenamefont {{Monney}},\ and\ \citenamefont {{St{\"a}hler}}}]{Mor2018inhibition}%
  \BibitemOpen
  \bibfield  {author} {\bibinfo {author} {\bibfnamefont {S.}~\bibnamefont {{Mor}}}, \bibinfo {author} {\bibfnamefont {M.}~\bibnamefont {{Herzog}}}, \bibinfo {author} {\bibfnamefont {J.}~\bibnamefont {{Noack}}}, \bibinfo {author} {\bibfnamefont {N.}~\bibnamefont {{Katayama}}}, \bibinfo {author} {\bibfnamefont {M.}~\bibnamefont {{Nohara}}}, \bibinfo {author} {\bibfnamefont {H.}~\bibnamefont {{Takagi}}}, \bibinfo {author} {\bibfnamefont {A.}~\bibnamefont {{Trunschke}}}, \bibinfo {author} {\bibfnamefont {T.}~\bibnamefont {{Mizokawa}}}, \bibinfo {author} {\bibfnamefont {C.}~\bibnamefont {{Monney}}},\ and\ \bibinfo {author} {\bibfnamefont {J.}~\bibnamefont {{St{\"a}hler}}},\ }\bibfield  {title} {\bibinfo {title} {{Inhibition of the photoinduced structural phase transition in the excitonic insulator Ta$_{2}$NiSe$_{5}$}},\ }\href {https://doi.org/10.1103/PhysRevB.97.115154} {\bibfield  {journal} {\bibinfo  {journal} {\prb}\ }\textbf {\bibinfo {volume} {97}},\ \bibinfo {pages} {115154} (\bibinfo {year}
  {2018})}\BibitemShut {NoStop}%
\bibitem [{\citenamefont {{Werdehausen}}\ \emph {et~al.}(2018{\natexlab{b}})\citenamefont {{Werdehausen}}, \citenamefont {{Takayama}}, \citenamefont {{Albrecht}}, \citenamefont {{Lu}}, \citenamefont {{Takagi}},\ and\ \citenamefont {{Kaiser}}}]{Werdehausen2018photo}%
  \BibitemOpen
  \bibfield  {author} {\bibinfo {author} {\bibfnamefont {D.}~\bibnamefont {{Werdehausen}}}, \bibinfo {author} {\bibfnamefont {T.}~\bibnamefont {{Takayama}}}, \bibinfo {author} {\bibfnamefont {G.}~\bibnamefont {{Albrecht}}}, \bibinfo {author} {\bibfnamefont {Y.}~\bibnamefont {{Lu}}}, \bibinfo {author} {\bibfnamefont {H.}~\bibnamefont {{Takagi}}},\ and\ \bibinfo {author} {\bibfnamefont {S.}~\bibnamefont {{Kaiser}}},\ }\bibfield  {title} {\bibinfo {title} {{Photo-excited dynamics in the excitonic insulator Ta$_{2}$NiSe$_{5}$}},\ }\href {https://doi.org/10.1088/1361-648X/aacd76} {\bibfield  {journal} {\bibinfo  {journal} {J. Phys. Condens.}\ }\textbf {\bibinfo {volume} {30}},\ \bibinfo {pages} {305602} (\bibinfo {year} {2018}{\natexlab{b}})}\BibitemShut {NoStop}%
\bibitem [{\citenamefont {{Ning}}\ \emph {et~al.}(2020)\citenamefont {{Ning}}, \citenamefont {{Mehio}}, \citenamefont {{Buchhold}}, \citenamefont {{Kurumaji}}, \citenamefont {{Refael}}, \citenamefont {{Checkelsky}},\ and\ \citenamefont {{Hsieh}}}]{ning2020signatures}%
  \BibitemOpen
  \bibfield  {author} {\bibinfo {author} {\bibfnamefont {H.}~\bibnamefont {{Ning}}}, \bibinfo {author} {\bibfnamefont {O.}~\bibnamefont {{Mehio}}}, \bibinfo {author} {\bibfnamefont {M.}~\bibnamefont {{Buchhold}}}, \bibinfo {author} {\bibfnamefont {T.}~\bibnamefont {{Kurumaji}}}, \bibinfo {author} {\bibfnamefont {G.}~\bibnamefont {{Refael}}}, \bibinfo {author} {\bibfnamefont {J.~G.}\ \bibnamefont {{Checkelsky}}},\ and\ \bibinfo {author} {\bibfnamefont {D.}~\bibnamefont {{Hsieh}}},\ }\bibfield  {title} {\bibinfo {title} {{Signatures of Ultrafast Reversal of Excitonic Order in Ta$_{2}$NiSe$_{5}$}},\ }\href {https://doi.org/10.1103/PhysRevLett.125.267602} {\bibfield  {journal} {\bibinfo  {journal} {\prl}\ }\textbf {\bibinfo {volume} {125}},\ \bibinfo {pages} {267602} (\bibinfo {year} {2020})}\BibitemShut {NoStop}%
\bibitem [{\citenamefont {{Gerber}}\ \emph {et~al.}(2017)\citenamefont {{Gerber}}, \citenamefont {{Yang}}, \citenamefont {{Zhu}}, \citenamefont {{Soifer}}, \citenamefont {{Sobota}}, \citenamefont {{Rebec}}, \citenamefont {{Lee}}, \citenamefont {{Jia}}, \citenamefont {{Moritz}}, \citenamefont {{Jia}}, \citenamefont {{Gauthier}}, \citenamefont {{Li}}, \citenamefont {{Leuenberger}}, \citenamefont {{Zhang}}, \citenamefont {{Chaix}}, \citenamefont {{Li}}, \citenamefont {{Jang}}, \citenamefont {{Lee}}, \citenamefont {{Yi}}, \citenamefont {{Dakovski}}, \citenamefont {{Song}}, \citenamefont {{Glownia}}, \citenamefont {{Nelson}}, \citenamefont {{Kim}}, \citenamefont {{Chuang}}, \citenamefont {{Hussain}}, \citenamefont {{Moore}}, \citenamefont {{Devereaux}}, \citenamefont {{Lee}}, \citenamefont {{Kirchmann}},\ and\ \citenamefont {{Shen}}}]{gerber2017femtosecond}%
  \BibitemOpen
  \bibfield  {author} {\bibinfo {author} {\bibfnamefont {S.}~\bibnamefont {{Gerber}}}, \bibinfo {author} {\bibfnamefont {S.~L.}\ \bibnamefont {{Yang}}}, \bibinfo {author} {\bibfnamefont {D.}~\bibnamefont {{Zhu}}}, \bibinfo {author} {\bibfnamefont {H.}~\bibnamefont {{Soifer}}}, \bibinfo {author} {\bibfnamefont {J.~A.}\ \bibnamefont {{Sobota}}}, \bibinfo {author} {\bibfnamefont {S.}~\bibnamefont {{Rebec}}}, \bibinfo {author} {\bibfnamefont {J.~J.}\ \bibnamefont {{Lee}}}, \bibinfo {author} {\bibfnamefont {T.}~\bibnamefont {{Jia}}}, \bibinfo {author} {\bibfnamefont {B.}~\bibnamefont {{Moritz}}}, \bibinfo {author} {\bibfnamefont {C.}~\bibnamefont {{Jia}}}, \bibinfo {author} {\bibfnamefont {A.}~\bibnamefont {{Gauthier}}}, \bibinfo {author} {\bibfnamefont {Y.}~\bibnamefont {{Li}}}, \bibinfo {author} {\bibfnamefont {D.}~\bibnamefont {{Leuenberger}}}, \bibinfo {author} {\bibfnamefont {Y.}~\bibnamefont {{Zhang}}}, \bibinfo {author} {\bibfnamefont {L.}~\bibnamefont {{Chaix}}}, \bibinfo {author} {\bibfnamefont
  {W.}~\bibnamefont {{Li}}}, \bibinfo {author} {\bibfnamefont {H.}~\bibnamefont {{Jang}}}, \bibinfo {author} {\bibfnamefont {J.~S.}\ \bibnamefont {{Lee}}}, \bibinfo {author} {\bibfnamefont {M.}~\bibnamefont {{Yi}}}, \bibinfo {author} {\bibfnamefont {G.~L.}\ \bibnamefont {{Dakovski}}}, \bibinfo {author} {\bibfnamefont {S.}~\bibnamefont {{Song}}}, \bibinfo {author} {\bibfnamefont {J.~M.}\ \bibnamefont {{Glownia}}}, \bibinfo {author} {\bibfnamefont {S.}~\bibnamefont {{Nelson}}}, \bibinfo {author} {\bibfnamefont {K.~W.}\ \bibnamefont {{Kim}}}, \bibinfo {author} {\bibfnamefont {Y.~D.}\ \bibnamefont {{Chuang}}}, \bibinfo {author} {\bibfnamefont {Z.}~\bibnamefont {{Hussain}}}, \bibinfo {author} {\bibfnamefont {R.~G.}\ \bibnamefont {{Moore}}}, \bibinfo {author} {\bibfnamefont {T.~P.}\ \bibnamefont {{Devereaux}}}, \bibinfo {author} {\bibfnamefont {W.~S.}\ \bibnamefont {{Lee}}}, \bibinfo {author} {\bibfnamefont {P.~S.}\ \bibnamefont {{Kirchmann}}},\ and\ \bibinfo {author} {\bibfnamefont {Z.~X.}\ \bibnamefont
  {{Shen}}},\ }\bibfield  {title} {\bibinfo {title} {{Femtosecond electron-phonon lock-in by photoemission and x-ray free-electron laser}},\ }\href {https://doi.org/10.1126/science.aak9946} {\bibfield  {journal} {\bibinfo  {journal} {Science}\ }\textbf {\bibinfo {volume} {357}},\ \bibinfo {pages} {71} (\bibinfo {year} {2017})}\BibitemShut {NoStop}%
\bibitem [{\citenamefont {{Huang}}\ \emph {et~al.}(2023)\citenamefont {{Huang}}, \citenamefont {{Querales-Flores}}, \citenamefont {{Teitelbaum}}, \citenamefont {{Cao}}, \citenamefont {{Henighan}}, \citenamefont {{Liu}}, \citenamefont {{Jiang}}, \citenamefont {{De la Pe{\~n}a}}, \citenamefont {{Krapivin}}, \citenamefont {{Haber}}, \citenamefont {{Sato}}, \citenamefont {{Chollet}}, \citenamefont {{Zhu}}, \citenamefont {{Katayama}}, \citenamefont {{Power}}, \citenamefont {{Allen}}, \citenamefont {{Rotundu}}, \citenamefont {{Bailey}}, \citenamefont {{Uher}}, \citenamefont {{Trigo}}, \citenamefont {{Kirchmann}}, \citenamefont {{Murray}}, \citenamefont {{Shen}}, \citenamefont {{Savi{\'c}}}, \citenamefont {{Fahy}}, \citenamefont {{Sobota}},\ and\ \citenamefont {{Reis}}}]{huang2023ultrafast}%
  \BibitemOpen
  \bibfield  {author} {\bibinfo {author} {\bibfnamefont {Y.}~\bibnamefont {{Huang}}}, \bibinfo {author} {\bibfnamefont {J.~D.}\ \bibnamefont {{Querales-Flores}}}, \bibinfo {author} {\bibfnamefont {S.~W.}\ \bibnamefont {{Teitelbaum}}}, \bibinfo {author} {\bibfnamefont {J.}~\bibnamefont {{Cao}}}, \bibinfo {author} {\bibfnamefont {T.}~\bibnamefont {{Henighan}}}, \bibinfo {author} {\bibfnamefont {H.}~\bibnamefont {{Liu}}}, \bibinfo {author} {\bibfnamefont {M.}~\bibnamefont {{Jiang}}}, \bibinfo {author} {\bibfnamefont {G.}~\bibnamefont {{De la Pe{\~n}a}}}, \bibinfo {author} {\bibfnamefont {V.}~\bibnamefont {{Krapivin}}}, \bibinfo {author} {\bibfnamefont {J.}~\bibnamefont {{Haber}}}, \bibinfo {author} {\bibfnamefont {T.}~\bibnamefont {{Sato}}}, \bibinfo {author} {\bibfnamefont {M.}~\bibnamefont {{Chollet}}}, \bibinfo {author} {\bibfnamefont {D.}~\bibnamefont {{Zhu}}}, \bibinfo {author} {\bibfnamefont {T.}~\bibnamefont {{Katayama}}}, \bibinfo {author} {\bibfnamefont {R.}~\bibnamefont {{Power}}}, \bibinfo {author}
  {\bibfnamefont {M.}~\bibnamefont {{Allen}}}, \bibinfo {author} {\bibfnamefont {C.~R.}\ \bibnamefont {{Rotundu}}}, \bibinfo {author} {\bibfnamefont {T.~P.}\ \bibnamefont {{Bailey}}}, \bibinfo {author} {\bibfnamefont {C.}~\bibnamefont {{Uher}}}, \bibinfo {author} {\bibfnamefont {M.}~\bibnamefont {{Trigo}}}, \bibinfo {author} {\bibfnamefont {P.~S.}\ \bibnamefont {{Kirchmann}}}, \bibinfo {author} {\bibfnamefont {{\'E}.~D.}\ \bibnamefont {{Murray}}}, \bibinfo {author} {\bibfnamefont {Z.-X.}\ \bibnamefont {{Shen}}}, \bibinfo {author} {\bibfnamefont {I.}~\bibnamefont {{Savi{\'c}}}}, \bibinfo {author} {\bibfnamefont {S.}~\bibnamefont {{Fahy}}}, \bibinfo {author} {\bibfnamefont {J.~A.}\ \bibnamefont {{Sobota}}},\ and\ \bibinfo {author} {\bibfnamefont {D.~A.}\ \bibnamefont {{Reis}}},\ }\bibfield  {title} {\bibinfo {title} {{Ultrafast Measurements of Mode-Specific Deformation Potentials of Bi$_{2}$Te$_{3}$ and Bi$_{2}$Se$_{3}$}},\ }\href {https://doi.org/10.1103/PhysRevX.13.041050} {\bibfield  {journal} {\bibinfo
  {journal} {Phys, Rev. X}\ }\textbf {\bibinfo {volume} {13}},\ \bibinfo {pages} {041050} (\bibinfo {year} {2023})}\BibitemShut {NoStop}%
\bibitem [{\citenamefont {{Chen}}\ \emph {et~al.}(2023{\natexlab{b}})\citenamefont {{Chen}}, \citenamefont {{Tang}}, \citenamefont {{Chen}}, \citenamefont {{Kang}}, \citenamefont {{Ding}}, \citenamefont {{Scott}}, \citenamefont {{Wang}}, \citenamefont {{Li}}, \citenamefont {{Ruff}}, \citenamefont {{Hashimoto}}, \citenamefont {{Lu}}, \citenamefont {{Jozwiak}}, \citenamefont {{Bostwick}}, \citenamefont {{Rotenberg}}, \citenamefont {{da Silva Neto}}, \citenamefont {{Birgeneau}}, \citenamefont {{Chen}}, \citenamefont {{Louie}}, \citenamefont {{Wang}},\ and\ \citenamefont {{He}}}]{chen2023anomalous}%
  \BibitemOpen
  \bibfield  {author} {\bibinfo {author} {\bibfnamefont {C.}~\bibnamefont {{Chen}}}, \bibinfo {author} {\bibfnamefont {W.}~\bibnamefont {{Tang}}}, \bibinfo {author} {\bibfnamefont {X.}~\bibnamefont {{Chen}}}, \bibinfo {author} {\bibfnamefont {Z.}~\bibnamefont {{Kang}}}, \bibinfo {author} {\bibfnamefont {S.}~\bibnamefont {{Ding}}}, \bibinfo {author} {\bibfnamefont {K.}~\bibnamefont {{Scott}}}, \bibinfo {author} {\bibfnamefont {S.}~\bibnamefont {{Wang}}}, \bibinfo {author} {\bibfnamefont {Z.}~\bibnamefont {{Li}}}, \bibinfo {author} {\bibfnamefont {J.~P.~C.}\ \bibnamefont {{Ruff}}}, \bibinfo {author} {\bibfnamefont {M.}~\bibnamefont {{Hashimoto}}}, \bibinfo {author} {\bibfnamefont {D.-H.}\ \bibnamefont {{Lu}}}, \bibinfo {author} {\bibfnamefont {C.}~\bibnamefont {{Jozwiak}}}, \bibinfo {author} {\bibfnamefont {A.}~\bibnamefont {{Bostwick}}}, \bibinfo {author} {\bibfnamefont {E.}~\bibnamefont {{Rotenberg}}}, \bibinfo {author} {\bibfnamefont {E.~H.}\ \bibnamefont {{da Silva Neto}}}, \bibinfo {author} {\bibfnamefont
  {R.~J.}\ \bibnamefont {{Birgeneau}}}, \bibinfo {author} {\bibfnamefont {Y.}~\bibnamefont {{Chen}}}, \bibinfo {author} {\bibfnamefont {S.~G.}\ \bibnamefont {{Louie}}}, \bibinfo {author} {\bibfnamefont {Y.}~\bibnamefont {{Wang}}},\ and\ \bibinfo {author} {\bibfnamefont {Y.}~\bibnamefont {{He}}},\ }\bibfield  {title} {\bibinfo {title} {{Anomalous excitonic phase diagram in band-gap-tuned Ta$_{2}$Ni(Se,S)$_{5}$}},\ }\href {https://doi.org/10.1038/s41467-023-43365-1} {\bibfield  {journal} {\bibinfo  {journal} {Nat. Commun.}\ }\textbf {\bibinfo {volume} {14}},\ \bibinfo {pages} {7512} (\bibinfo {year} {2023}{\natexlab{b}})}\BibitemShut {NoStop}%
\end{thebibliography}
\end{document}